\documentclass[aps,prx, onecolumn,showpacs,superscript address,preprint]{revtex4-1}
\usepackage{amsmath}
\usepackage{upgreek}
\usepackage{graphicx}
\usepackage[euler]{textgreek}
\usepackage[export]{adjustbox}
\usepackage{float}
\usepackage{physics}
\usepackage{mathrsfs}
\usepackage{bm}
\usepackage{caption}
\usepackage{textcomp}
\usepackage{natbib}
\usepackage{gensymb}
\usepackage{amsfonts} 
\usepackage{hyperref} 
\usepackage{xcolor}

\begin{document}
\captionsetup[figure]{labelsep=period,name={FIG.}, format={plain},justification={raggedright}}
\title{Roles of band gap and Kane electronic dispersion in the THz-frequency nonlinear optical response in HgCdTe}

\author{Davide Soranzio}
\affiliation{Institute for Quantum Electronics, Eidgenössische Technische Hochschule (ETH) Zürich, CH-8093 Zurich, Switzerland}
\author{Elsa Abreu}
\affiliation{Institute for Quantum Electronics, Eidgenössische Technische Hochschule (ETH) Zürich, CH-8093 Zurich, Switzerland}
\author{Sarah Houver}
\affiliation{Université Paris Cité, CNRS, Matériaux et Phénomènes Quantiques, F-75013, Paris, France}
\author{Janine D{\"o}ssegger}
\affiliation{Institute for Quantum Electronics, Eidgenössische Technische Hochschule (ETH) Zürich, CH-8093 Zurich, Switzerland}
\author{Matteo Savoini}
\affiliation{Institute for Quantum Electronics, Eidgenössische Technische Hochschule (ETH) Zürich, CH-8093 Zurich, Switzerland}
\author{Frédéric Teppe}
\affiliation{Laboratoire Charles Coulomb, UMR CNRS 5221, University of Montpellier, Montpellier 34095, France}
\author{Sergey Krishtopenko}
\affiliation{Laboratoire Charles Coulomb, UMR CNRS 5221, University of Montpellier, Montpellier 34095, France}
\author{Nikolay N. Mikhailov}
\affiliation{A.V. Rzhanov Institute of Semiconductor Physics, Siberian Branch of Russian Academy of Sciences, Novosibirsk, Russia}
\affiliation{Novosibirsk State University, Novosibirsk, Russia}
\author{Sergey A. Dvoretsky}
\affiliation{A.V. Rzhanov Institute of Semiconductor Physics, Siberian Branch of Russian Academy of Sciences, Novosibirsk, Russia}
\affiliation{Tomsk State University, Tomsk, Russia}
\author{Steven L. Johnson}
\affiliation{Institute for Quantum Electronics, Eidgenössische Technische Hochschule (ETH) Zürich, CH-8093 Zurich, Switzerland}
\email[e-mail:]{johnson@phys.ethz.ch}

\date{\today}

\begin{abstract}
Materials with linear electronic dispersion often feature high carrier mobilities and unusually strong nonlinear optical interactions. In this work, we investigate the THz nonlinear dynamics of one such material, HgCdTe, with an electronic band dispersion heavily dependent on both temperature and stoichiometry. We show how the band gap, carrier concentration and band shape together determine the nonlinear response of the system. At low temperatures, carrier generation from Zener tunneling dominates the nonlinear response with a reduction in the overall transmission. At room temperature, quasi-ballistic electronic dynamics drive the largest observed nonlinear optical interactions, leading to a transmission increase. Our results demonstrate the sensitivity of these nonlinear optical properties of narrow-gap materials to small changes in the electronic dispersion and carrier concentration.
\end{abstract}

\maketitle

\section{Introduction}
Materials such as graphene and topological semimetals with pseudo-relativistic band dispersion in the vicinity of the Fermi level have been reported to host a wide variety of interesting phenomena \cite{Wang2015,Armitage2018}. Some of these effects arise from the strong impact of the gapless electronic band dispersion and massless quasiparticles on the optical properties of the materials, which leads to strong nonlinear light-matter interactions such as carrier multiplication and high-order harmonic generation \cite{Tani2012,Bowlan2014,Baudisch2018}. The relative importance of key parameters, such as the carrier concentration and small changes in the band gap of such materials in determining the nonlinear response, has at present not been well explored or understood, especially for interactions in the far infrared range (300~GHz--20~THz).  

One example of a material with nearly-linear electronic dispersion under certain conditions is bulk mercury cadmium telluride (MCT, Hg\textsubscript{$1-x$}Cd\textsubscript{$x$}Te). MCT is an alloy of HgTe and CdTe commonly found in infrared detectors \cite{Norton2002}\cite{Lei2015}. HgTe is a semimetal, while CdTe is a semiconductor with a band gap of 1.5~eV. The small lattice mismatch between the two compounds enables the synthesis of alloys with any value of doping $x$ between 0 and 1 \cite{Dornhaus1976}.
Near the $\Gamma$ point, the electronic band structure of MCT can be described using a $k\cdot p$ model \cite{Kane1980}. The band gap in bulk MCT can be tuned to some extent and even eliminated using temperature \cite{Teppe2016}, biaxial strain \cite{Brune2011}, hydrostatic pressure \cite{Szola2022} and randomly distributed impurities \cite{Krishtopenko2022}.
At regions of the phase diagram at the boundary between semimetallic and semiconducting phases, the electronic band structure shows a 3D linear dispersion of pseudo-relativistic quasiparticles called Kane fermions \cite{Orlita2014,Teppe2016}. These quasiparticles share characteristics of ultrarelativistic Dirac fermions \cite{Krishtopenko2022_2}.
Unlike Weyl and Dirac semimetals, in MCT the band crossing is not topology- or symmetry-protected, although a \begin{math}\mathbb{Z}_{\mathrm{2}}\end{math} invariant can be defined to differentiate the phases.

In this work, we study the dynamics of a MCT film with $x\approx 0.175$ using broadband 0.5--5 THz pulses, as a function of temperature and amplitude of the pulsed electric field. At this composition, MCT is a semiconductor with a band structure that changes strongly with temperature: as the temperature decreases, the band gap becomes narrower and the conduction band dispersion nearly linear. The change in band gap with temperature keeps the thermally activated carriers at relatively low concentrations across the whole temperature range \cite{Teppe2016}. This causes the equilibrium chemical potential to be close to the bottom of the conduction band for all temperatures.
We use 1D and 2D broadband THz time-domain spectroscopies (TDS) \cite{Kuehn_2009} to investigate the low-frequency electrodynamics of this material in a transmission geometry.

We interpret our results by performing simulations using the finite-difference time-domain (FDTD) method \cite{Yee1966}. We model carrier dynamics including the ballistic response and evaluating two different carrier generation mechanisms namely Zener tunneling \cite{Kane1960} and impact ionization \cite{Dick1972}.
Finally, we compare the results of the experiments and simulations and discuss how band gap, carrier concentration and band shape determine the nonlinear response of the system. MCT serves as a model system for the interpretation of these phenomena, which can then be generalized to more complex narrow-band gap semiconductors.

\section{Methods}
\subsection{Experimental}
The sample is a MCT heterostructure grown on a 400-\textmu{}m-thick semi-insulating (SI) GaAs substrate with lateral dimensions $5\times 3$ mm$^2$ by the procedure described by Dvoretsky et al. \cite{Dvoretsky2019}.
Thin layers of CdTe, between 5-7 \textmu m thick, and ZnTe, approximately 30 nm thick, act as buffer layers between the substrate and the 3.2-\textmu{m}-thick (013) MCT film with doping $x\approx0.175$, accompanied by two few-hundred-nm-thick MCT layers at its surfaces along the stacking direction with a monotonic increase of the Cd content to avoid stress and defects caused by a mismatch in the lattice parameters. More details regarding the sample structure can be found in Orlita et al. \cite{Orlita2014} and in the Supplemental Material, Section I \cite{supplementary}. For our measurements, the sample was mounted on a copper plate in front of a 2.7-mm circular hole and inserted into a helium cryostat.

The THz response was measured in two separate setups.
Low-field, with peak values $\approx$4 kV/cm, 1D-TDS data was collected using a setup described previously by Suter et al. \cite{Suter2022}, focusing the THz beam to a diameter of $450$~\textmu{}m at the sample position and using electro-optic detection of the THz waveform. A Ti:Sapphire regenerative amplifier produces 50 fs full-width-at-half-maximum (FWHM) 800 nm pulses at a 250 kHz repetition rate. These pulses are split into two paths. One path directs the beam onto a spintronic emitter, generating quasi-single-cycle THz pulses which are then sent through the sample. The other beam path samples the transmitted electric field using electro-optic sampling in a 300-\textmu m-thick (110) GaP crystal.

A schematic of the high-field 2D-TDS setup is provided in the Supplemental Information, Section II \cite{supplementary}.
In short, a laser system based on a Ti:Sapphire regenerative amplifier produces femtosecond pulses (800 nm, 100 fs FWHM, 1 kHz). A beamsplitter divides these pulses, which are then directed into a pair of three-stage optical parametric amplifiers (OPAs). The signal output from each OPA is individually tuneable to wavelengths near 1.5 \textmu m. Mechanical choppers inserted into the two signal outputs create pulse trains at submultiples of the 1 kHz laser repetition rate: one at 500 Hz (beam 1) and the other at 250 Hz (beam 2). To create THz-frequency pulses, each beam is then focused into a crystal of 4-N,N-dimethylamino-4’-N’-methyl-stilbazolium 2,4,6-trimethylbenzenesulfonate (DSTMS). Optical rectification of the pulses creates broadband THz radiation centered at approximately 2 THz \cite{Vicario2015}. For each beam a pair of wire-grid polarizers controls the electric field amplitude and polarization of the THz pulses. The pulses are then combined, directed toward the sample using plane mirrors and focused to a diameter of approximately 450 \textmu{}m (FWHM) with off-axis parabolic mirrors. A series of off-axis parabolic mirrors then refocuses the transmitted THz into a GaP electro-optic sampling crystal in combination with a small part of the 800 nm beam split before the OPAs. The parameters of the GaP crystal are the same as in the 1D measurements.
To estimate the peak fields and waveforms of the incident pulses, we remove the sample and measure the fields for each pulse train individually.
The absolute field amplitudes are estimated using the relations found in Hirori et al. \cite{Hirori2011_2}.
To reduce the contribution of the quadratic Kerr response, we inserted calibrated Si filters after the sample to reduce the field amplitude arriving at the GaP crystal until we observed a linear dependence of the field amplitude on additional attenuation. 

The different chopping rates for the two beams allow to record quasi-simultaneously the transmitted signal for four distinct field combinations: 
both THz pulses ($E$\textsubscript{t}$(t)$), only one of the two pulses ($E$\textsubscript{1}$(t)$, $E$\textsubscript{2}$(t)$) and no THz pulse ($E$\textsubscript{b}$(t)$, background).
For the measurements performed without the sample in place, we define $E$\textsubscript{k,i}$(t) = E$\textsubscript{k}$(t) - E$\textsubscript{b}$(t)$ as the `input' waveform, where k=t,1,2. We assume that the functions $E_{\mathrm{k,i}}$ are a good approximation of the incident electric field waveforms on the sample for the respective cases. For measurements with the sample in place, we define $E$\textsubscript{k,o}$(t) = E$\textsubscript{k}$(t) - E$\textsubscript{b}$(t)$ as the `output' waveform, where again the subscript k indicates which set of pulses is incident on the sample.
Both beams enter the sample at normal incidence, with parallel electric field polarizations along the crystallographic [$\bar{3}3\bar{1}$] direction, as verified through Laue x-ray diffraction. 
The THz beam paths were maintained in a low-humidity environment ($<$2\% RH) to minimize absorption by water vapor.

For each transmitted single-pulse waveform we define the time centroid as 
\begin{equation}
t_{\mathrm{k}} = \int |E_{\mathrm{k,o}}(t)|^2 t dt / \int|E_{\mathrm{k,o}}(t)|^2 dt
\end{equation}
where k=1,2 \cite{Reimann_2007}.
To describe the relative timing of the two pulses and the electro-optic sampling beam we also define two time coordinates: the excitation delay $t_{\mathrm{ex}} = t_{\mathrm{2}}-t_{\mathrm{1}}$, which is the difference between the time centroids, and the detection delay $t_{\mathrm{del}} = t-t_{\mathrm{2}}$, where \(t\) is the time probed by electro-optic sampling. 

To isolate nonlinear effects, we define the `nonlinear' signal $E_{\mathrm{nl,o}}=E_{\mathrm{t,o}}-E_{\mathrm{1,o}}-E_{\mathrm{2,o}}$ and the `nonlinear-probe' signal $E_{\mathrm{nlp,o}}=E_{\mathrm{t,o}}-E_{\mathrm{1,o}}$, which depend on the temporal superposition of the individual fields.
The typical waveforms of the THz pulses employed in the 1D- and 2D-TDS experiments can be found in the Supplemental Information, Sections III and IV \cite{supplementary}.

\subsection{FDTD Simulations}

The numerical simulations were carried out based on the discretized version of Maxwell's equations proposed by Yee \cite{Yee1966}, as described by Huber \cite{Huber2017} following the scheme from Rumpf \cite{Rumpf2022}. All fields are assumed to depend on only one spatial coordinate $z$, where $z=0$ defines the boundary between air and the sample.
The local electric displacement field at a time $t$ can be written as 
\begin{equation}
\mathbf{D}(z,t)=\epsilon_\mathrm{0}\varepsilon_\mathrm{\infty}(z)\mathbf{E}(z,t)+\mathbf{P}_{\mathrm{e}}(z,t)+\mathbf{P}_{\mathrm{ph}}(z,t)
\end{equation}
where $\epsilon_\mathrm{0}$ is the vacuum permittivity, $\varepsilon_\mathrm{\infty}(z)$ is the high-frequency relative permittivity of the medium at $z$, $\mathbf{E}(z,t)$ is the electric field, $\mathbf{P}_{\mathrm{e}}(z,t)$ the electronic polarization density and $\mathbf{P}_{\mathrm{ph}}(z,t)$ the polarization density from low-frequency infrared-active vibrations. Here we assume $\mathbf{P}_{\mathrm{e}}(z,t)$ includes only contributions from free carriers, and that contributions from other electronic transitions or high-frequency vibrational modes are included in $\varepsilon_\mathrm{\infty}(z)$ and are time-independent. In our simulations we include the MCT layer, the two buffer layers (CdTe and ZnTe) and the GaAs substrate. We assume that $\mathbf{P}_{\mathrm{e}}(z,t)=0$ for all materials except for the MCT layer and $\mathbf{P}_{\mathrm{ph}}(z,t)=0$ for all materials except for the MCT and CdTe layers.
The sample is modeled as a 1D succession of layers with different dielectric constants based on the composition profile reported in \cite{Orlita2014,Dvoretsky2019}. A detailed description can be found in the Supplemental Information, Section I \cite{supplementary}.

To model the electronic polarization density $\mathbf{P}_{\mathrm{e}}(z,t)$, we assume that the conduction band electrons with density $n(z,t)$ behave as a quasi-ballistic wavepacket with an average wavevector $\mathbf{k}(z,t)$~\cite{Houver2019}.
The dynamics of $\mathbf{P}_{\mathrm{e}}(z,t)$ are then determined by
\begin{equation}
\frac{\partial \mathbf{P}_{\mathrm{e}}(z,t)}{\partial t} = -e n(z,t) \mathbf{v}[\mathbf{k}(z,t) ]\label{eq:dPedt}
\end{equation}
where $e$ is the fundamental charge, 
\begin{equation}
\mathbf{v}(\mathbf{k}) = \frac{1}{\hbar} \nabla_{\mathbf{k}} \mathcal{E}(\mathbf{k})\label{eq:groupv}
\end{equation}
is the wavepacket group velocity, $\hbar=h/2\pi$ with $h$ Planck's constant and $\mathcal{E}(\mathbf{k})$ is the energy of the conduction band at $\mathbf{k}$. 

Regarding $\mathcal{E}(\mathbf{k})$, we use the electronic band structure for Hg$_{0.825}$Cd$_{0.175}$Te in an isotropic approximation, shown to be effective to describe its magnetoabsorption \cite{Teppe2016}. We estimate based on the data of Teppe et al. \cite{Teppe2016} that the band gap varies between 9 and 138 meV over the temperature range 13 K to 300 K, approaching gapless linear dispersion at low temperatures. In such conditions, the band structure near the Fermi level can be described using an isotropic simplified Kane model \cite{Kane1980} valid in the vicinity of the bottom of the $\Gamma$ valley. The conduction band dispersion is expressed by the quasi-relativistic relation \cite{Teppe2016}
\begin{equation}
\mathcal{E}_{\mathrm{c}}(k)=-\mathcal{E}_{\mathrm{g}}/2+\sqrt{ m^{*2}_{\mathrm{o}}\tilde c^4+\hbar^2k^2\tilde c^2}
 \label{eq1}
\end{equation}
where $\mathcal{E}_{\mathrm{g}}$ the electronic band gap, $m^{*}_{\mathrm{o}}$ is the effective mass at the bottom of the conduction band, $\tilde c$ is the asymptotic velocity of Kane fermions (weakly-dependent on temperature) and $k$ is the wavevector magnitude. Since $m^{*}_{\mathrm{o}}\tilde c^2\approx \mathcal{E}_{\mathrm{g}}/2$ \cite{Teppe2016}, $\mathcal{E}_{\mathrm{c}}=0$ corresponds to the bottom of the conduction band.

We now need to determine the wavevector of the wavepacket $\mathbf{k}(z,t)$.
In the ballistic regime, this is given by the equation of motion \cite{Yu2017}
\begin{equation}
\hbar\frac{\partial\mathbf{k}(z,t)}{\partial t}+\gamma_{\mathrm{P}}\hbar \mathbf{k}(z,t)=-e\mathbf{E}(z,t)
\label{eqball}
\end{equation}
\noindent
where $\gamma_{\mathrm{P}}$ is an electron scattering term. We assume that initially $\mathbf{k}(z,t) = 0$.

The carrier density $n(z,t)$ is initially set at a value based on the temperature-dependent measurements of Teppe et al.~\cite{Teppe2016} (see also Section III in \cite{supplementary}). For the highest electric fields used in our experiments, however, we expect that the density of carriers may increase~\cite{Yu2017,Hoffmann2009}. In the simulations we tested two possible carrier generation mechanisms, namely Zener tunneling \cite{Kane1960,Sanari2018} and impact ionization \cite{Dick1972,Hoffmann2009}. 

The Zener interband tunneling effect has been observed in several narrow-gap semiconductors in the presence of a static electric field \cite{Zener1934,Kane1960} and it has been previously reported for Hg$_{1-x}$Cd$_x$Te diodes \cite{Bhan1994}. Since the time steps in our simulation are much smaller than the field oscillation period, we can approximate the propagation of our THz pulses as a succession of constant-field steps. The instantaneous rate of carrier generation is then \cite{Kane1960} 
\begin{equation}
\frac{\partial n(z,t)}{\partial t}\bigg|_{\mathrm{Zener}} =\frac{e^2E(z,t)^2}{18\pi\hbar^2}\left ( \frac{m_{\mathrm{\mu}}}{\mathcal{E}_{\mathrm{g}}}\right )^{1/2}\mathrm{exp}\left(\frac{-\pi m_{\mathrm{\mu}}^{1/2}\mathcal{E}_{\mathrm{g}}^{3/2} }{2\hbar e \left | E(z,t)\right |} \right)
 \label{eq2}
\end{equation}
where $E(z,t)$ is the local electric field magnitude, $m_\mu=m^{*}_{\mathrm{o}} m^{\mathrm{*}}_{\mathrm{hh}}/(m^{*}_{\mathrm{o}}+m^{\mathrm{*}}_{\mathrm{hh}})$ is the reduced mass, and $m^{\mathrm{*}}_{\mathrm{hh}}=0.55\cdot m_{\mathrm{e}}$ is the effective mass of the heavy-hole bands at the $\Gamma$ point with $m_{\mathrm{e}}$ the electron rest mass (from \cite{Kinch2004}, Chapter 7).

Impact ionization can be modelled based on the estimate of probability per unit time for impact ionization in MCT (Eq. 7.9, from Kinch \cite{Kinch2004})
\begin{multline}
\mathcal{P}(\mathcal{E}_\mathrm{n})=\frac{3.2\cdot10^7Ve^4m^{*}_{\mathrm{o}}[F_1F_2]^2_{\mathrm{o}}\rho_\mathrm{v}}{\varepsilon^2_\mathrm{0}h^3\epsilon^2_\mathrm{0}}\\
\int^{\mathcal{E}_\mathrm{n}}_{1}dx
\frac{(a^2+1)x^2}{a^2+x^2} \frac{\left\{[2(\mathcal{E}_\mathrm{n}-x)+1]^2-1\right\}^{1/2}[2(\mathcal{E}_\mathrm{n}-x)+1] [2x-1]\left\{[2x-1]^2-1\right\}^{1/2}}{[(\lambda/k_{\mathrm{g}})^2+x^2]^2}
\end{multline}
where $\mathcal{E}_\mathrm{n}=\mathcal{E}_{\mathrm{c}}/\mathcal{E}_{\mathrm{g}}$ is the ratio between the electron energy with respect to the bottom of the conduction band and the band gap, $V$ is the unit cell volume \cite{Reine2005}, $a=\sqrt{10}$ is a numerical value chosen to adjust the overlap integral $[F_1F_2]^2_{\mathrm{o}}$ (values reported in Kinch \cite{Kinch2004}), $\rho_\mathrm{v}=(\mathcal{E}_{\mathrm{g}} m_{\mathrm{o}}^{*} m^{\mathrm{*}2}_{\mathrm{hh}}/2)^{1/2}8\pi/h^3$ is the density of states in the heavy-hole band \cite{Kinch2004}, $\varepsilon_\mathrm{0}$ is the low-frequency ($\omega\to0$) relative permeability from Baars and Sorger \cite{Baars1972}, $\epsilon_\mathrm{0}$ is the vacuum permeability, $\lambda$ is the screening parameter of the Coulomb interaction, assumed to be negligible ($\lambda=0$) during impact ionization \cite{Kinch2004}, and $k_{\mathrm{g}}=\sqrt{8m^*_{\mathrm{o}}\mathcal{E}_{\mathrm{g}}\pi^2/h^2}$ \cite{Kinch2004}.
In order to take into account the momentum change of the electron wavepacket due to the scattering event, we added to the left-hand side of Eq. \ref{eqball} an effective damping force $-\mathbf{F}_{\mathrm{imp}}(\mathbf{k})$ following the procedure described by Biasco et al. \cite{Biasco2022}, such that
\begin{equation}
\left.\mathbf{F}_{\mathrm{imp}}=\hbar\frac{\partial \mathbf{k}}{\partial t}\right|_\textrm{imp}=-\frac{\hbar}{2}\frac{(1+2\alpha \mathcal{E}_{\mathrm{c}})\left(\mathcal{E}_{\mathrm{c}}-\mathcal{E}_{\mathrm{th}}\right ) \mathcal{P}(\mathcal{E}_\mathrm{n})}{(1+\alpha \mathcal{E}_{\mathrm{c}})\mathcal{E}_{\mathrm{c}}} \theta\left ( \mathcal{E}_{\mathrm{c}} -\mathcal{E}_{\mathrm{th}} \right )\mathbf{k}
\end{equation}
where $\alpha=1/\mathcal{E}_{\mathrm{g}}$ \cite{Zawadzki1974} and we assumed $\mathcal{E}_{\mathrm{th}}\approx \mathcal{E}_{\mathrm{g}}$ \cite{Hoffmann2009}.
We can then write
\begin{equation}
\left.\frac{\partial n(z,t)}{\partial t}\right|_\textrm{imp} = \mathcal{P}[\mathcal{E}_\mathrm{n}(z,t)]n(z,t).\label{eq:dndtimp}
\end{equation}
In principle, we could now define a total rate of change for the carrier density by adding the contributions from Eqs.~\ref{eq2} and~\ref{eq:dndtimp}. In practice, we instead simulated scenarios where only one carrier generation mechanism was active.  

Now we turn to the third term on the right-hand-side of Eq.~\ref{eq1}, the phonon-polariton contribution to the polarization density. We approximate MCT as having two phonon-polariton modes in our frequency range, so that $\mathbf{P}_{\mathrm{ph}}(z,t)=\mathbf{P}_1(z,t)+\mathbf{P}_2(z,t)$ and
\begin{equation}
\frac{\partial^2\mathbf{P}_\mathrm{m}(z,t)}{\partial t^2}+\Gamma_{\mathrm{m}}\frac{\partial\mathbf{P}_\mathrm{m}(z,t)}{\partial t}+\omega^2_\mathrm{TO,m}\mathbf{P}_\mathrm{m}(z,t)=\varepsilon_\mathrm{\infty}(z)\left(\omega^2_\mathrm{LO,m}-\omega^2_\mathrm{TO,m}\right)\mathbf{E}(z,t)
\end{equation}
where $\nu_\mathrm{LO,m}=\omega_\mathrm{LO,m}/2\pi$ and $\nu_\mathrm{TO,m}=\omega_\mathrm{TO,m}/2\pi$ are the longitudinal and transversal frequencies and $\gamma_{\mathrm{m}}$ is the damping rate of the m-th mode. We assume that all polarization densities are initially zero. Further details can be found in the Supplemental Material, Section I.

\section{Results}
\subsection{1D-TDS}
\begin{figure}[t]
\includegraphics[width=\columnwidth]{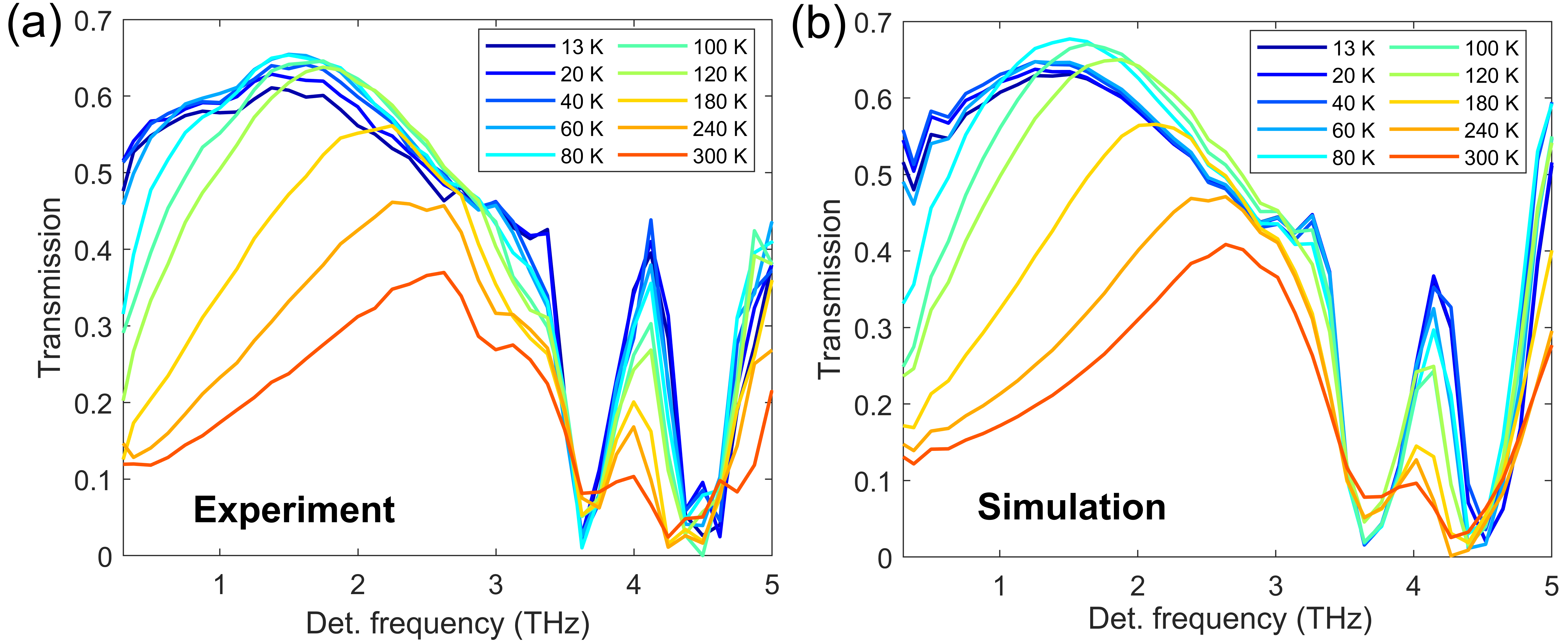}

\caption{Low field ($\approx$4 kV/cm) (a) experimental and (b) FDTD-simulated 1D-TDS transmission spectra of Hg$_{0.825}$Cd$_{0.175}$Te at different temperatures.}

\label{fig1}
\end{figure}

We first performed low-field 1D-TDS measurements at different temperatures to understand the linear response of the material.
In Fig. \ref{fig1}(a), the transmission spectra of the sample are shown for a range of temperatures.
As the sample is warmed up, a gradual increase of the carrier concentration leads to a change of the left slope of the spectrum together with an overall decrease of the transmission. 
A shallow dip in the transmission appears close to 3.0 THz for higher temperatures, which has been observed in previous work \cite{Grynberg1974,Polit2010,Kozyrev2011}.
The two dips around 3.6 THz and 4.4 THz have also been observed earlier and attributed to Hg-Te and Cd-Te vibrations, where the largest contributions come from modes which are both Raman- and IR-active \cite{Amirtharaj1990,Kozyrev1998}. 

Fig. \ref{fig1}(b) shows the results of our simulation of the 1D-TDS, after adjusting the electron carrier concentration and scattering rate together with the longitudinal and transversal frequencies and scattering rate of the phonon-polariton terms. The complete set of parameters can be found in the Supplemental Material, Section III \cite{supplementary}. The simulations were performed including the contribution of Zener tunneling to carrier generation.
For the lowest temperatures the best-fit parameters for the carrier concentration are close to the values reported by Teppe et al. \cite{Teppe2016}, while at higher temperatures we observed that the best fit was found for concentrations of a factor approximately 2 lower.

The decrease in transmission as temperature grows is well-captured by the simulations. The two transmission minima above 3 THz are accurately reproduced in the simulations, where they arise from vibrational resonances.
The shallow dip at approximately 3 THz seen at high temperatures is not captured by the simulations. This feature has been resolved before
but has an unclear origin~\cite{Grynberg1974,Polit2010,Kozyrev2011}. It may be related to additional vibrational resonances that we do not include in the model.

\begin{figure}[h!]
\includegraphics[width=\columnwidth]{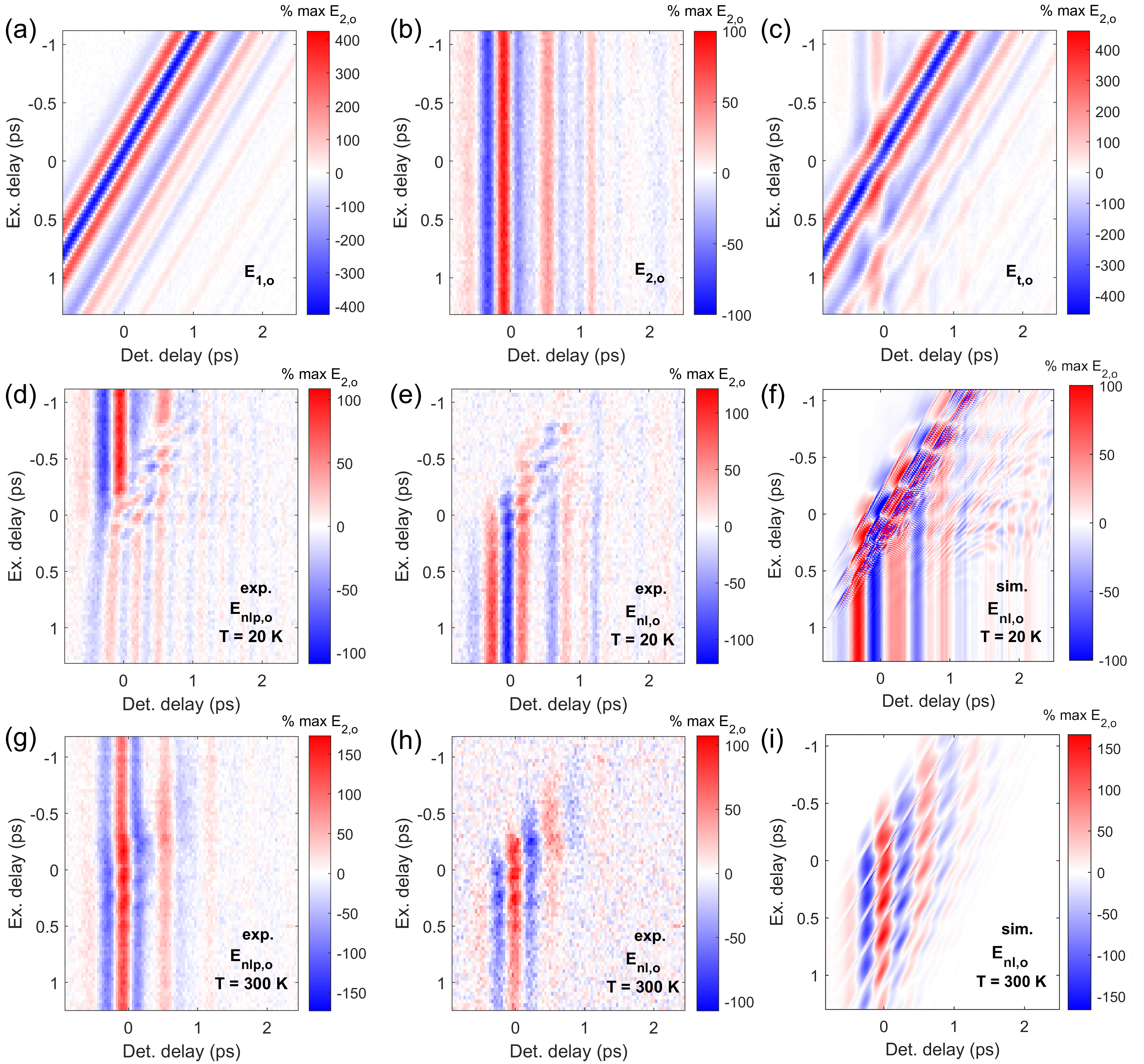}

\caption{2D-TDS maps. (a-c) Measurements of 
(a) $E_{\mathrm{1,o}}$, (b) $E_{\mathrm{2,o}}$ and (c) $E_{\mathrm{t,o}}$ acquired at $T=20$ K for peak fields $E_{\mathrm{1,i}}=16.9$~kV/cm and $E_{\mathrm{2,i}}=7.7$~kV/cm. (d) Experimental nonlinear probe $E_{\mathrm{nlp,o}}$ and (e) nonlinear $E_{\mathrm{nl,o}}$ signals recorded at $T=20$ K for peak fields $E_{\mathrm{1,i}}$=169.4 kV/cm and $E_{\mathrm{2,i}}=7.7$~kV/cm. Panel (e) can be directly compared with the FDTD-simulated $E_{\mathrm{nl,o}}$ in panel (f), using the same incident field amplitudes. (g-h) The experimental nonlinear probe $E_{\mathrm{nlp,o}}$ and experimental nonlinear $E_{\mathrm{nl,o}}$ signals measured at $T=300$~K and the same pulse field strengths as in (d-e). (i) The simulated $E_{\mathrm{nl,o}}$ at $T=300$~K, using the same incident field amplitudes as (h).}

 \label{fig2}
 \end{figure}


\begin{figure}[h]
\includegraphics[width=\textwidth]{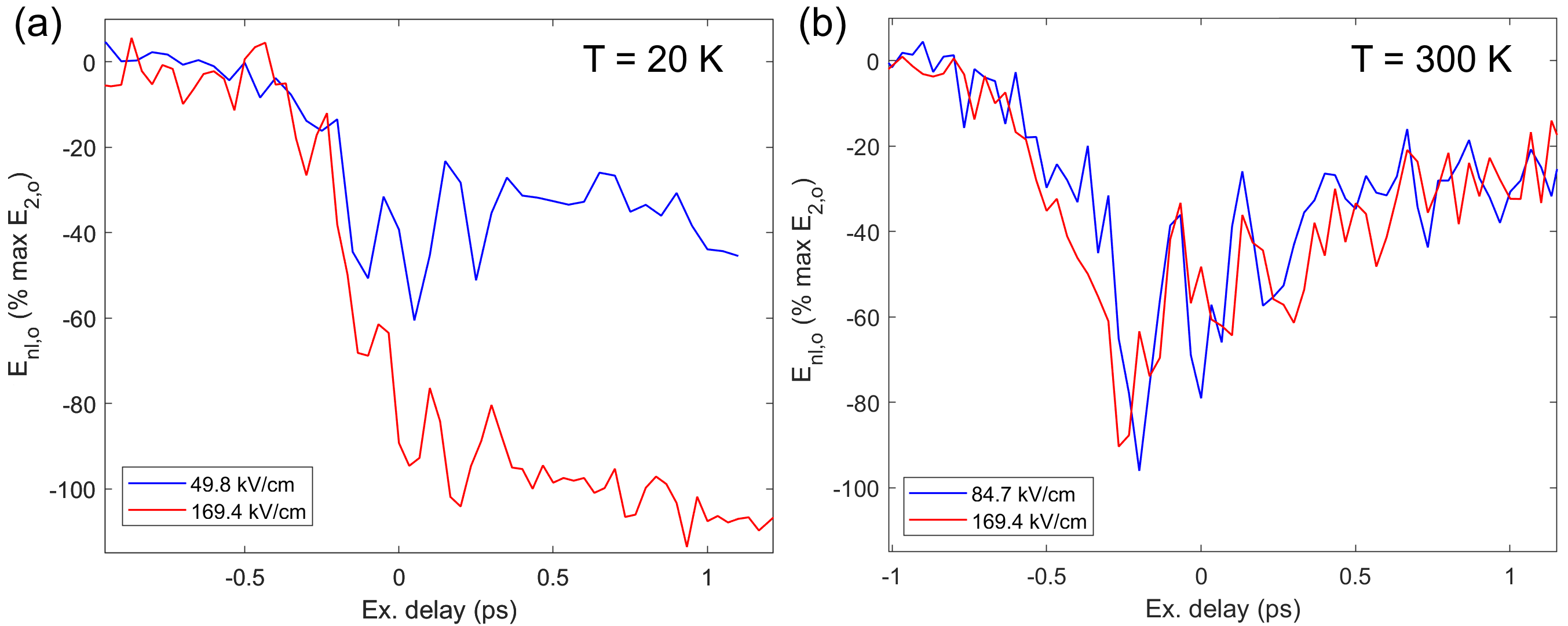}

\caption{Time-resolved excitation-delay profiles $E_{\mathrm{nl,o}}$ for peak $E_{\mathrm{2,i}}=7.7$~kV/cm and two different $E_{\mathrm{1,i}}$ field amplitudes at constant detection delay averaged at (a) $T=20$~K ($t_{\mathrm{del}}= -0.06$~ps, in a $\pm50$ fs window) and (b) $T=300$~K ($t_{\mathrm{del}}= 0.23$~ps in a $\pm33$ fs window).
}
 \label{fig3}
\end{figure}

\subsection{2D-TDS}

In Fig. \ref{fig2} we report 2D-TDS maps at two different temperatures. To better understand the nature of the nonlinear signal, the three experimental datasets $E_{\mathrm{1,o}}$, $E_{\mathrm{2,o}}$ and $E_{\mathrm{t,o}}$ are shown in panels (a)-(c) for T=20 K and peak $E_{\mathrm{1,i}}$=16.9 kV/cm and $E_{\mathrm{2,i}}$=7.7 kV/cm; using two pulses with comparable amplitudes here allows to distinguish their contributions to $E_{\mathrm{t,o}}$. Panels (d) and (g) show the nonlinear probe $E_{\mathrm{nlp,o}}$ at 20 K and 300 K, respectively, with peak $E_{\mathrm{1,i}}$=169.4 kV/cm and $E_{\mathrm{2,i}}$=7.7 kV/cm as a function of excitation delay $t_{\mathrm{ex}}$ and detection delay $t_{\mathrm{del}}$. The nonlinear probe can be roughly interpreted as the transmitted $E_{2}$ field with respect to the arrival time of the $E_{1}$ field. At 20 K, $E_{\mathrm{nlp,o}}$ first oscillates, then quickly decreases by almost 80\% after the arrival of $E_{1}$, remaining unchanged for excitation delays of several picoseconds (see Supplemental Material, Section V \cite{supplementary}). In contrast, at 300 K we observe more extended oscillations around the pulse overlap region together with a slow increase in $E_{\mathrm{nlp,o}}$ that relaxes within a couple of picoseconds. 
A complementary view on the phenomenon is given by the nonlinear signal $E_{\mathrm{nl,o}}$ shown in panels (e) and (h) at 20 K and 300 K, respectively. The evolution of the response as a function of the time delay between the two pulses can also be visualized taking constant detection-delay profiles as reported in Fig.~\ref{fig3} for two different peak amplitudes of $E_{\mathrm{1,i}}$ at $T=20$ K (panel (a)) and $T=300$ K (panel (b)).
In the $T=20$ K comparison, we observe that the modulation period decreases for higher peak field, in contrast to what observed for $T=300$ K. We have also taken 2D nonlinearity maps over somewhat longer excitation delay ranges for a selection of intermediate temperatures and input fields (Section V, ~\cite{supplementary}). For the higher peak fields $E_{\mathrm{1,i}}$ at $T=300$~K, these data show a slow decrease in $E_{\mathrm{nlp,o}}$ that emerges on a time scale that shortens with increasing field strength, down to approximately 5~ps at 169.4~kV/cm. These changes in $E_{\mathrm{nlp,o}}$ appear similar to the long-excitation-delay behavior of $E_{\mathrm{nlp,o}}$ seen at $T=20$ K at lower fields, but are much smaller in magnitude. We also observe changes in the single-pulse transmission as a function of field strength, which are summarized in Appendix A.

For our simulations of the 2D nonlinear response, we use the electronic and vibrational parameters determined from adapting the 1D simulations to match our data, without further adjustment. We initially performed simulations assuming only Zener tunneling as a mechanism for carrier generation. 
The results shown in Fig. \ref{fig2}(f),(i) capture the clear distinction between the $T=20$ K and $T=300$ K cases that we observe in the corresponding experimental maps (panels (e), (h)). At $T=20$~K the simulations well reproduce the approximate magnitude of the nonlinear response as well as the behavior at long excitation delays (above 1 ps). In the region where the two pulses overlap in time, the simulation shows some high frequency modulations in the nonlinear signal along a diagonal direction, which does not appear in the experimental maps. We comment on this in Appendix A.
For $T=300$~K the simulation reproduces the approximate overall magnitude and modulation seen in the experimental data, but has a significantly higher modulation visibility.

\begin{figure}[t]
\includegraphics[width=\textwidth]{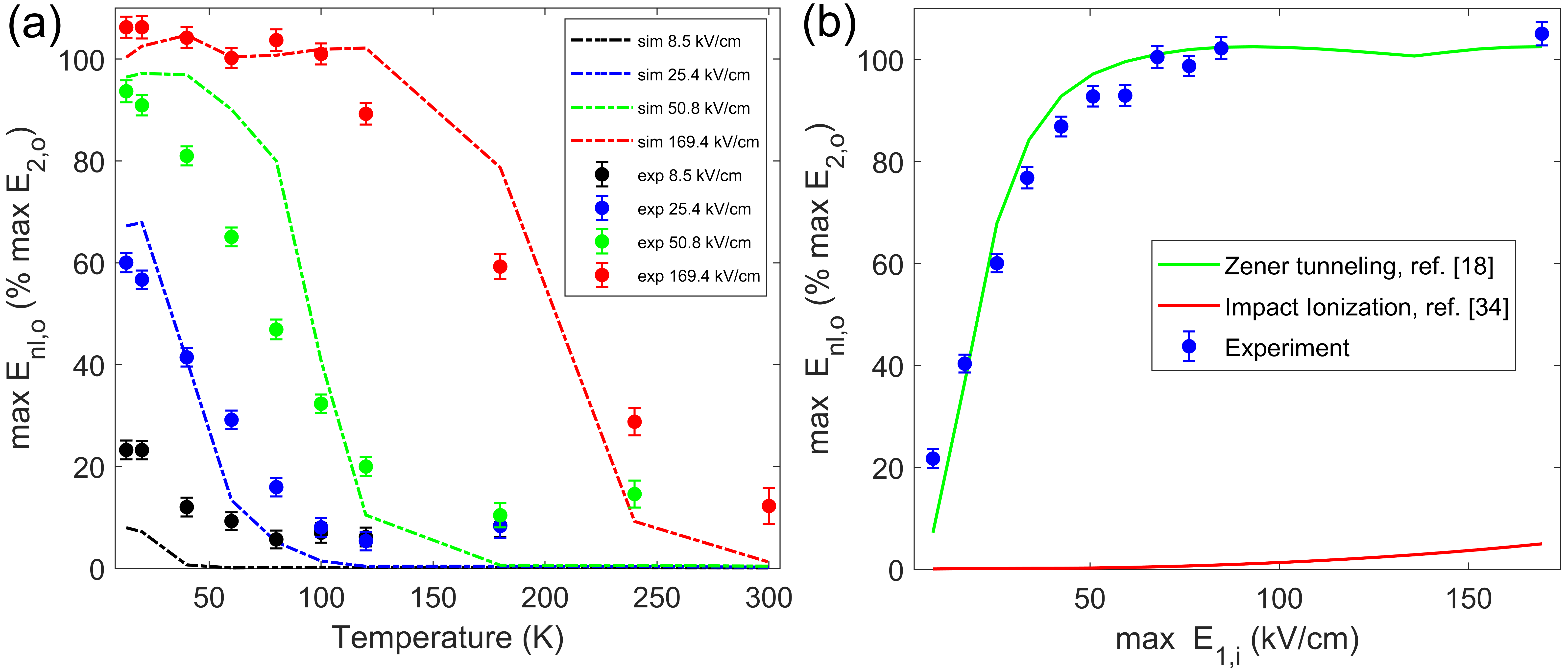}

\caption{Peak amplitudes of the nonlinear effects observed at $t_{\mathrm{ex}}$=2.54 ps (averaged around a $\pm 67$ fs range). (a) Experimental and simulated temperature dependence at selected fields (b) Experimental and simulated field dependence at T=20 K employing different carrier generation mechanisms.}

\label{fig4}
\end{figure}

One interesting feature of the 2D maps is the change in behavior at long excitation delays as the temperature increases: at $T=20$~K we see a drop in the transmission of pulse 2 that becomes nearly independent of excitation delay, whereas at $T=300$~K we see a relaxation on a time scale of about 1 ps. 
To study this aspect of the data in more detail, we acquired constant-excitation-delay traces at $t_{\mathrm{ex}}=2.54$ ps, averaged over a $\pm 67$ fs range, as a function of both temperature and peak field $E_{\mathrm{1,i}}$. In Fig.~\ref{fig4}(a) we show the maximum value of the measured nonlinear signal $E_{\mathrm{nl,o}}$ over our detection delay range as a function of temperature for different peak fields of pulse 1. For comparison we also show the results of our simulation. For all peak field values, we see that the nonlinear signal increases with decreasing temperature, with an apparent threshold temperature where the effects become quite strong. For the highest fields we observe a saturation in the nonlinear response that is also reproduced quite accurately by the simulations. 
Fig.~\ref{fig4}(b) 
shows the maximum nonlinear signal as a function of peak field at $T=20$~K. Here we also show the results of the simulations with Zener tunneling and impact ionization individually as the only carrier generation mechanism. We see here that while the simulation of Zener tunneling fairly accurately describes the magnitude and saturation of the nonlinear response at high fields, the simulation of impact ionization severely underestimates the magnitude of the nonlinearity and does not reproduce the saturation behavior seen in the experiment.

\begin{figure}[h]
\includegraphics[width=\textwidth]{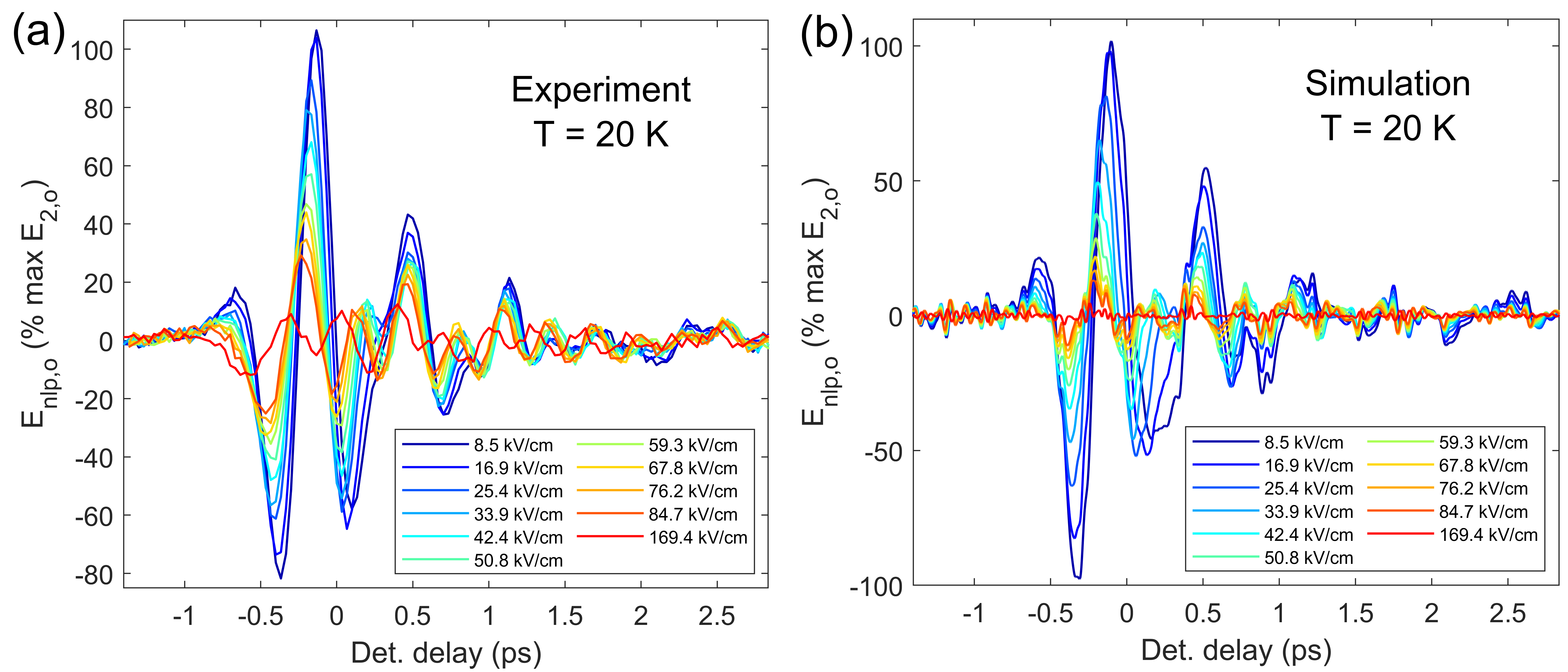}

\caption{(a) Experimental and (b) FDTD-simulated peak $E_{\mathrm{1,i}}$ field dependence of $E_{\mathrm{nlp,o}}$ acquired at $t_{\mathrm{ex}}$=2.54 ps averaged around a $\pm 67$ fs range at T=20 K.}

 \label{fig5}
\end{figure}

In addition to examining the behavior of the maximum of the nonlinear effect at long excitation delays, we can also study the dependence of the nonlinear probe signal on the detection delay. 
Fig. \ref{fig5} shows the full $E_{\mathrm{nlp,o}}$ trace as a function of detection delay associated with the data summarized Fig. \ref{fig4}(b). Panel (a) shows the gradual reduction of the pulse transmission as the $E_{i,1}$ field amplitude grows. This change is accompanied by a modification of the relative spectral weight of the transmitted frequency components, leading to an approximately 3 THz oscillation for high fields. 
The simulated counterpart (using Zener tunneling) is reported in panel (b). The transmission of the $E_{\mathrm{2,i}}$ field decreases as the $E_{\mathrm{1,i}}$ field increases and the best quantitative correspondence is found for intermediate fields. As with the experimental data, at high $E_{\mathrm{1,i}}$ field amplitudes the spectral content of the transmitted pulse shifts towards higher frequency.

Further experimental results on the parity of the nonlinear effects, the experimental uncertainties and the SI-GaAs spectrum are reported in the Supplemental Material, Sections VI-VIII, \cite{supplementary}. Additional TDS simulations of the NL effects using only impact ionization as the carrier generation mechanism as a function of temperature and $E_1$ peak field amplitude are shown in the Supplemental Material, Section IX, \cite{supplementary}.

\section{Discussion}

In the 1D-TDS measurements, the overall decrease in transmission with increasing temperature seen in Fig.~\ref{fig1} has been explained in earlier work as the result of an increase in the density of thermally-activated carriers, a conclusion that is consistent with our simulation results (\cite{Teppe2016,Grynberg1974} and Section III,   \cite{supplementary}). 

Counter to this trend, a small increase in the transmission was detected from 13 to 60 K between 1 and 2 THz. This is reproduced by the simulations, and appears to be connected with a reduction of the electronic damping parameter $\gamma_{\mathrm{P}}$. 
This may indicate a decrease of the scattering pathways due to the opening of the band gap ~\cite{Chang2004}.

Now we turn to the discussion of the nonlinear 2D data shown in Figs.~\ref{fig2} and~\ref{fig3}. 
Around the region of the temporal superposition between the two pulses, we observe oscillations in the nonlinear signal which, from the comparison between experiments and simulations, we attribute to quasi-ballistic motion of the carriers. 
At low temperature, the nonlinear response rapidly reaches a plateau mainly linked to carrier generation effects. The larger carrier concentration causes a reduced transmission of pulse 2 after the arrival of pulse 1. From a comparison between the two carrier generation models shown in Fig. \ref{fig4}(b), we identify the predominant mechanism as Zener tunneling \cite{Kane1960}.
Besides reducing the transmission of pulse 2, carrier generation also changes the spectral content of the transmitted field. As seen in Fig.~\ref{fig5}, at higher peak fields of pulse 1, the average period of the transmitted portion of pulse 2 is markedly smaller, suggesting that the higher carrier concentration filters out low frequencies preferentially. This is consistent with the data of Fig.~\ref{fig1}, where we see that as carrier concentration increases with temperature, lower frequencies are more attenuated. This is even evident in the times when the pulses overlap, as shown in Fig.~\ref{fig3}(a) where we show the low- and high-field nonlinear signal at $T = 20$ K as a function of excitation delay. Here we see that for higher peak field values the periodicity of the modulation decreases, suggesting that here too lower frequency components are attenuated more strongly as carriers are generated.

For higher sample temperatures, carrier generation processes are strongly reduced due to the increase of the band gap \cite{Teppe2016} (Fig. \ref{fig4}(a),(b)). 
As a result, the nonlinear signal for short excitation delay ranges is determined by the ballistic response and confined to the times when the pulses overlap. In contrast to the nonlinear signal at low temperatures, at high temperatures the sign of the nonlinearity indicates a transient \textit{increase} in the transmission of pulse 2.

\begin{figure}[h]
\includegraphics[width=\textwidth]{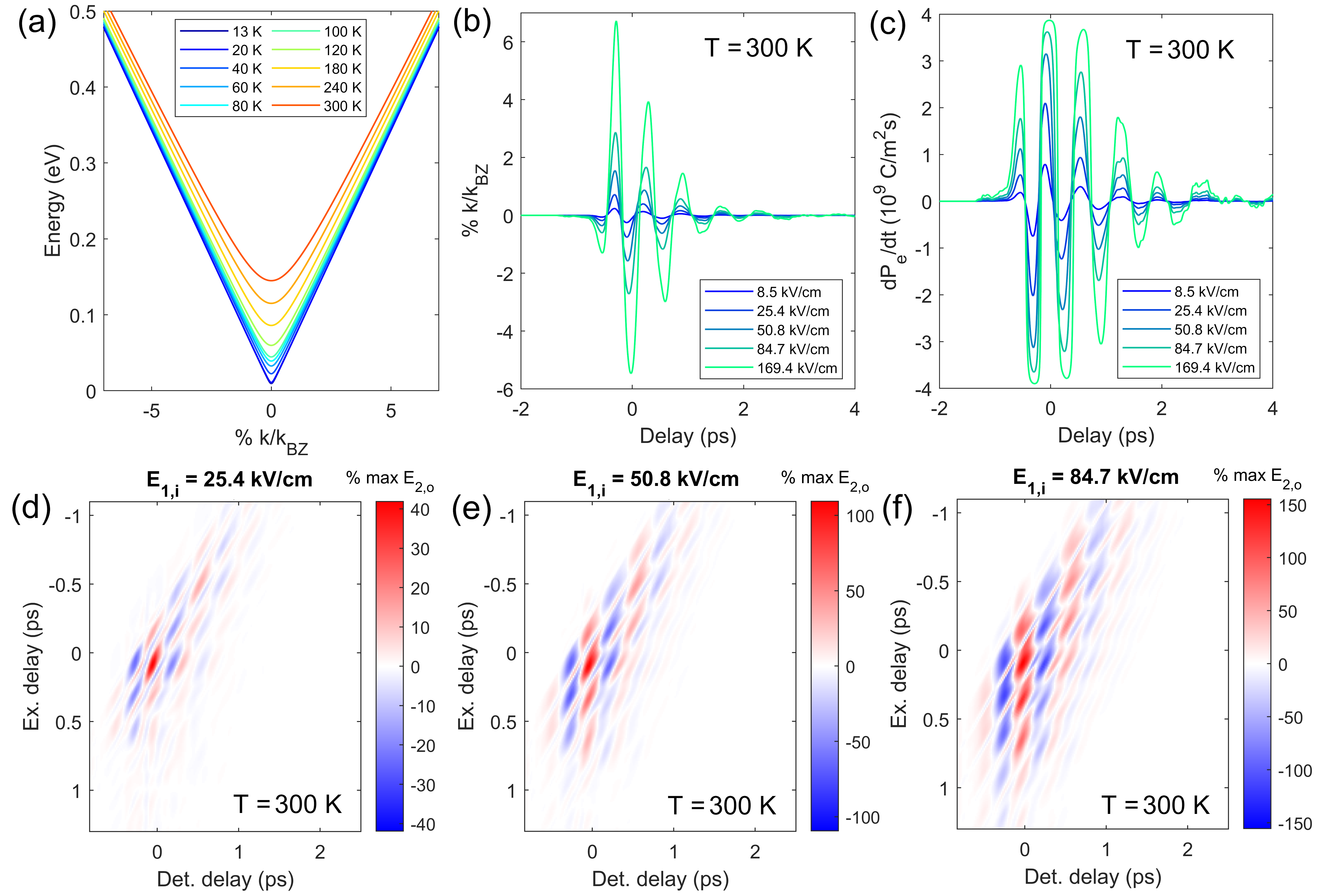}

\caption{(a) Conduction band using the parameters used for the FDTD simulations. 
(b) Simulated reciprocal space excursion of the electron wavepacket using $E_1$ single pulses of different field amplitudes. (c) Simulated temporal derivative of the electronic polarization for different field amplitudes. For the last two panels, the results refer to the first simulation cell of the Hg$_{0.825}$Cd$_{0.175}$Te layer.
(d)-(f) Simulated field dependence of $E_{\mathrm{nl,o}}$ at T=300 K and peak $E_{\mathrm{2,i}}$=7.7 kV/cm.}

 \label{fig6}
\end{figure}

The source of the nonlinearities for the ballistic response comes from the differences in the dynamics of the electronic polarization among the different combinations of the two pulses. In fact, the rate of change in the polarization density, described by Eq.~\ref{eq:dPedt}, is proportional to the velocity $\mathbf{v}[\mathbf{k}(t)]$, which strongly depends on the history of the time-dependent interaction between electron wavepacket and electric field up to the instant $t$ (Eqs. \ref{eq:groupv}, \ref{eqball}). Thus the ballistic nonlinear signal can occur in temporal regions where the evolution $\mathbf{v}[\mathbf{k}(t)]$ is significantly altered by the simultaneous application of the two pulses rather than it being simply the sum of the trajectories in reciprocal space given by the two individual pulses.
Moreover, the rate described by Eq.~\ref{eq:dPedt} is proportional to the carrier density $n$, favouring the higher temperatures which possess a larger carrier concentration.
There are, however, other changes that occur with an increase in temperature that affect the ballistic nonlinearities. These variations are a consequence of modifications to the conduction band dispersion as the temperature increases (Fig. \ref{fig6}(a), \cite{Orlita2014,Teppe2016}).
According to Eq.~\ref{eqball}, an increase in the amplitude of the local electric field $\mathbf{E}$ will translate into a proportional increase in the amplitude of oscillations in the momentum space. In Fig. ~\ref{fig6}(b) we show a simulation of the dynamics of the wavepacket wavevector $k$ as a function of pulse 1 peak amplitude that confirms this relationship. The level of the nonlinear response, however, is determined not by the change of the momentum $k$, but by the variations in the time derivative of the electronic polarization density $P_{\mathrm{e}}$, which is, according to Eq.~\ref{eq:dPedt}, proportional to the group velocity of the wavepacket. This leads to a saturation of the electronic polarization density for high driving fields, as shown in Fig. ~\ref{fig6}(c) due to the peculiar, asymptotically-linear, conduction dispersion (Eq. \ref{eq1}). As a result, ballistic nonlinearities with close magnitude occur over a wider range of excitation delays than the temporal profile of pulse 1 might suggest, as shown by the simulated 2D-TDS nonlinear signal in Fig. \ref{fig6}(d)-(f), obtained by varying the peak amplitude of the incident field $E_{\mathrm{1,i}}$.

As shown in Fig. \ref{fig6}(a), the saturation of the group velocity occurs much closer to the $\Gamma$ point as the temperature decreases. Therefore one would expect saturation effects in ballistic nonlinearities to occur at lower field amplitudes at low temperature. In reality, in such conditions the nonlinear response is heavily influenced by carrier generation processes. This limits the visibility of these features at low temperatures. In absence of carrier generation, we would expect indeed a very strong dependence of the ballistic nonlinearities on the band structure, as shown though further simulations in the Supplemental Information, Section X ~\cite{supplementary}.

The slower, approximately 1 ps time-scale dynamics present at high temperature (Fig. \ref{fig2}(h), Fig. \ref{fig3}(b)), is semi-quantitatively reproduced by our simulations (Fig. \ref{fig2}(i)) by considering ballistic transport. A similar increase in the transmission could be in principle caused by direct or assisted intervalley scattering towards the neighbouring L and X valleys, depending on how far the electrons travel from the bottom of the $\Gamma$ valley. This was shown to occur in similar zinc-blende compounds \cite{Ho2014,Ašmontas2020}.
In this scenario, the increase in transmission is determined by the fact that electron mobility is much lower in the neighbouring valleys due to the larger effective mass, leading to a reduction of the equilibrium plasma frequency. In the literature, tight-binding band structures calculated for different alloy compositions \cite{Becker2010} suggest that a high ($>$1 eV) barrier electron energy is required, while density functional theory simulations for HgTe suggest a lower threshold ($\approx$0.5 eV) \cite{Feng2010}.
From our simulations, the peak energy acquired by the electrons at 300 K under pulses with peak field amplitude of 169.4 kV/cm is not sufficient ($<$0.5 eV) to escape from the $\Gamma$ valley. This suggests that a strong contribution by intervalley scattering is unlikely in our case.

At sufficiently long time scales, we expect the electrons to return gradually to the bottom of the $\Gamma$ valley, thermalize and recombine with holes of the valence bands mainly through Shockley-Read-Hall, Auger recombination or radiative decay \cite{Becker2010,Aleshkin2021}. This and other slower changes in the conduction band occupation may explain the slow dynamics in $E_{\mathrm{nlp,o}}$ as a function of excitation delay seen at high peak fields (Section V, ~\cite{supplementary}).
Excitation and relaxation phenomena over such a time scale have a key role in determining the response and design of very-long wavelength detectors based on the material, where the control of generation and scattering processes is central \cite{Li2022,Wehmeier2023}.

While capturing many of the main features of the nonlinear response, our simulations have some limitations.
One arises from treating the entire carrier population as a single electron wavepacket, instead of an ensemble via, for example, a Monte Carlo treatment \cite{Brennan1991}. This is a practical trade-off: ensemble treatments are much more computationally intensive, and would be difficult to include in our simulations. An ensemble-based simulation would broaden the velocity distribution of the particles. This might cause a smearing-out of the oscillations seen for times corresponding to the overlap of the pulses closer to the experimental results (Figs. \ref{fig2},(h),(i) and \ref{fig3}(b)).

As noted above, based on a comparison of our simulations with the experiment, we conclude that Zener tunneling is the dominant carrier generation process. 
One approach to discuss the validity of our model of Zener tunneling is to consider the Keldysh parameter $\gamma_{\mathrm{K}}=2\pi\nu_\mathrm{0}\sqrt{2m_{\mathrm{e}}\mathcal{E}_{\mathrm{g}}}/(eE)$, where $\nu_\mathrm{0}$ is the light frequency, $m_{\mathrm{e}}$ is the electron mass, $\mathcal{E}_{\mathrm{g}}$ is the energy gap, $e$ the fundamental charge and $E$ the electric field magnitude \cite{Keldysh1965,Lange2014,Sanari2018}. Strictly speaking our treatment of Zener tunneling assumes $\gamma_{\mathrm{K}}\ll1$. In MCT at $T=13$~K (where the gap is within the bandwidth of the THz pulse), the Keldysh parameter becomes unitary around 40 kV/cm for $\nu_\mathrm{0} =2$~THz. This suggests that we may be approaching the limits of validity for the Zener tunneling model at the lowest temperatures using low field amplitudes. 
Extensions and corrections to the original Zener derivation have been proposed by several authors~\cite{Zener1934,Kane1960,Fritsche1966,Bhan1994,Chakraborty1985} that may improve agreement. For example, the extension proposed by Bhan and Gopal~\cite{Bhan1994} effectively reduces the contribution of carrier generation at high fields and could improve the field dependence of the nonlinear signal. A revised model could also in principle take into account the effects of having carriers already in the conduction band.

Despite our conclusion that we do not see evidence of significant impact ionization in our experiment, it has been reported to occur in quantum wells containing HgCdTe both for much longer, approximately 100 ns, THz pulses and in experiments using a static electric field bias~\cite{Kinch2004,Rees2010,Hubmann2019}. We believe that in these works impact ionization is more significant due to the much longer pulse (or field application) duration and more moderate electric fields, which combine to dramatically enhance the influence of impact ionization relative to that of Zener tunneling. A qualitative simulation of this is presented in the Supplemental Material, Section XI, ~\cite{supplementary}.

Lastly, we consider the possibility that the CdTe and ZnTe buffer layers and GaAs substrate may contribute to the observed nonlinearity.
These materials are larger-gap semiconductors (1.5 eV for CdTe, 2.3 eV for ZnTe and 1.5 eV for GaAs) compared to our MCT sample (\cite{Kasap2006} and are nominally semi-insulating \cite{Dvoretsky2019}.
While reports have shown that it is possible to induce nonlinearities in \textit{n}-doped GaAs at field amplitudes comparable to the ones used in our experiment, on SI-GaAs this has shown to be feasible only through a large local field enhancement using metallic structures \cite{Hirori2011,Fan2013,Kim2023}. We therefore rule out a major contribution to the nonlinear signal coming from those layers. Furthermore, the transmission spectrum of SI-GaAs is only weakly temperature-dependent in our frequency window as reported in the Supplemental Material, Section VIII \cite{supplementary}.

\section{Conclusions}
We have shown how the band gap, carrier concentration and band shape intertwine to determine the carrier dynamics in the narrow-gap semiconductor Hg$_{0.825}$Cd$_{0.175}$Te and how they heavily affect the nonlinear response of the material to transient THz fields. At low temperature, where the band gap is small, the initially ballistic dynamics leads to carrier generation mainly via Zener tunneling, which dominates the nonlinear response. At high temperature, under a higher carrier concentration, larger band gap and more parabolic-like band shape, a pronounced nonlinear ballistic response confined around the temporal overlap region of the two pulses was observed while carrier generation is heavily suppressed. 
Moreover, these phenomena influence the carrier dynamics over tens of picoseconds, which in turn impact the response and design of very-long wavelength detectors based on the material. The nonlinear response to ultrafast pulses is therefore seen to depend strongly on experimental parameters such as temperature, carrier concentration and field amplitude. While the present work is based on MCT, the key concepts presented here can be easily extended to analogous narrow-gap semiconductors.

\section*{Acknowledgments}
This work was supported by the Swiss National Science Foundation under project 200020\_192337.
E.A. acknowledges funding support from the Swiss National Science Foundation through Ambizione Grant PZ00P2\_179691 and and through Starting Grant TMSGI2\_211211.
This work was also supported by the CNRS through IRP ‘TeraMIR’ and by the "ANR" for Equipex+ Hybat project (ANR-21 -ESRE-0026).
High-performance computing resources were obtained from ETH Zürich (Euler cluster).

\appendix

\section{Intrapulse dynamics}

\begin{figure}[h]
\includegraphics[width=\textwidth]{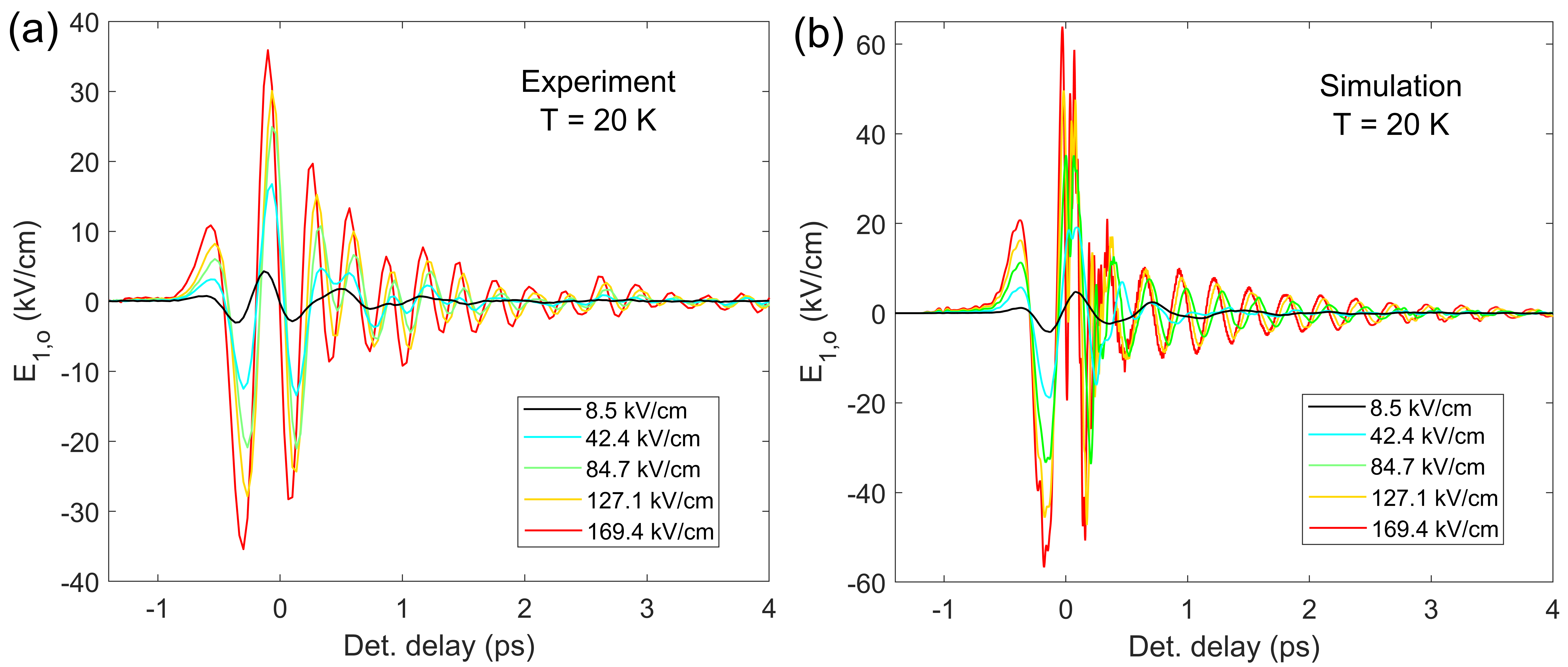}

\caption{(a) Experimental and (b) simulated peak $E_{\mathrm{1,i}}$ field amplitude dependence of the transmitted $E_{\mathrm{1,o}}$ pulse through MCT multilayer at T=20 K.}

 \label{fig7}
\end{figure}

In Fig. \ref{fig7} (a), we report the experimental time-domain trace of the transmitted $E_{\mathrm{1,o}}$ pulse. Here the input field amplitude was varied by rotating the first of a pair of wire-grid polarizers, which causes a frequency-independent attenuation. For comparison, panel (b) shows the corresponding simulated data.

The spectral shape of the transmitted $E_{\mathrm{1,o}}$ field shows clear differences on varying the input field amplitude.
Similar effects occur in semiconductors like \textit{n}-doped GaAs, where it has been attributed to self-phase modulation from carrier generation during the pulse~\cite{Turchinovich2012}. The first oscillations of the terahertz pulse are enough to lead to a considerable increase in the carrier population which in turn influence the transmission of the following part of the pulse. This is why for the first one or two oscillations the difference between the waveforms is well-approximated as a simple rescaling of the amplitude, whereas the rest of the transmitted pulse appears to have a higher frequency, due to the spectral filtering effect of the increased carrier density discussed in the main text. The presence of sudden spikes in correspondence of the highest field amplitudes, around $t_{\mathrm{del}}$ $\approx$ 0 ps in panel (b), is connected to the limitations of our FDTD algorithm and the single electron wavepacket approximation instead of a electronic ensemble.

\nocite{Schall2001,Moore1996,Hansen1982,Katayama2018,Fornies-Marquina1997}
\clearpage

\bibliography{biblio}

\begin{thebibliography}{68}
\bibitem [{sup()}]{supplementary}%
  \BibitemOpen
  \href@noop {} {\enquote {\bibinfo {title} {See {S}upplemental {M}aterial at
  [url will be inserted by publisher] for additional details on the sample,
  experimental data and simulation results, which includes refs. [64-68]},}\
  }\BibitemShut {NoStop}%
\end{thebibliography}

\end{document}


\captionsetup[figure]{labelsep=period,name={FIG.}, format={plain},justification={raggedright}}
\title{Supplemental Material on Roles of band gap and Kane electronic dispersion in the THz-frequency nonlinear optical response in HgCdTe}

\author{Davide Soranzio}
\affiliation{Institute for Quantum Electronics, Eidgenössische Technische Hochschule (ETH) Zürich, CH-8093 Zurich, Switzerland}
\author{Elsa Abreu}
\affiliation{Institute for Quantum Electronics, Eidgenössische Technische Hochschule (ETH) Zürich, CH-8093 Zurich, Switzerland}
\author{Sarah Houver}
\affiliation{Université Paris Cité, CNRS, Matériaux et Phénomènes Quantiques, F-75013, Paris, France}
\author{Janine D{\"o}ssegger}
\affiliation{Institute for Quantum Electronics, Eidgenössische Technische Hochschule (ETH) Zürich, CH-8093 Zurich, Switzerland}
\author{Matteo Savoini}
\affiliation{Institute for Quantum Electronics, Eidgenössische Technische Hochschule (ETH) Zürich, CH-8093 Zurich, Switzerland}
\author{Frédéric Teppe}
\affiliation{Laboratoire Charles Coulomb, UMR CNRS 5221, University of Montpellier, Montpellier 34095, France}
\author{Sergey Krishtopenko }
\affiliation{Laboratoire Charles Coulomb, UMR CNRS 5221, University of Montpellier, Montpellier 34095, France}
\author{Nikolay N. Mikhailov}
\affiliation{A.V. Rzhanov Institute of Semiconductor Physics, Siberian Branch of Russian Academy of Sciences, Novosibirsk, Russia}
\affiliation{Novosibirsk State University, Novosibirsk, Russia}
\author{Sergey A. Dvoretsky}
\affiliation{A.V. Rzhanov Institute of Semiconductor Physics, Siberian Branch of Russian Academy of Sciences, Novosibirsk, Russia}
\affiliation{Tomsk State University, Tomsk, Russia}
\author{Steven L. Johnson}
\affiliation{Institute for Quantum Electronics, Eidgenössische Technische Hochschule (ETH) Zürich, CH-8093 Zurich, Switzerland}

\date{\today}

\maketitle
\section{Further details on the numerical simulations}

For our numerical simulations we considered the layered nature of the sample,  as reported in \cite{Dvoretsky2019,Orlita2014}.  As a function of the THz beam propagation coordinate \(z\) (where +\(\hat{z}\) is the direction of propagation), the sample is modelled as follows:
\begin{itemize}
\item[] $z < 0$:  Vacuum
\item[] $0 < z < z_{\mathrm{1}} = d$:  Hg$_{1-x}$Cd$_{x}$Te (MCT) with a depth-dependent \(x(z)\) as reported by Orlita et al. \cite{Orlita2014}, and with a nominal total thickness $d = 5.3$ \textmu m 
\item[] $z_{\mathrm{1}} < z < z_{\mathrm{2}} = z_{\mathrm{1}} + d^\prime$: CdTe buffer layer with thickness \(d^\prime\)
\item[] $z_{\mathrm{2}} < z < z_{\mathrm{3}} = z_{\mathrm{2}} + d^{\prime\prime}$:  ZnTe buffer layer, also with thickness \(d^{\prime\prime}\)
\item[] $z_{\mathrm{3}} < z < z_{\mathrm{4}} = z_{\mathrm{3}} + d^{\prime\prime\prime}$:  GaAs with thickness $d^{\prime\prime\prime} = 400$ \textmu m
\item[] $z > z_{\mathrm{4}}$:  Vacuum
\end{itemize}
The value for $d^\prime$ was estimated to be between 5-7 \textmu m. For the simulations we decided to use the mean value, \textit{i.e.}, 6 \textmu m. The thickness of the ZnTe buffer layer $d^{\prime\prime}$ was approximately 30 nm. This is, however, very small compared the other thicknesses.  It is also small in comparison to the wavelengths of the THz light used:  for a frequency of 2 THz the index of refraction is approximately 3.2 in ZnTe~\cite{Schall2001}, giving a wavelength of approximately 46 \textmu m.  To make the simulations computationally efficient, we employ a step size of \(\Delta z = 160\) nm when evaluating the propagation of the fields, and we set \(d^{\prime\prime} = 160\)~nm so that exactly one step is evaluated in the ZnTe buffer layer.  In view of the very small thicknesses we do not expect this to have an appreciable effect on the validity of the simulation in our parameter range.

We estimate the value of \(d^{\prime\prime\prime}\)  by the experimentally measured time delay between the THz pulse transmitted by the sample and the pulse that is reflected from the \(z=0\) and \(z=z_{\mathrm{4}}\) interfaces, neglecting the thin layers and assuming an effective (frequency independent) index of refraction in GaAs of 3.59~\cite{Moore1996}.

As described in the main text, Eq.~2 gives the material-dependent relation between \(\mathbf{D}(z,t)\) and \(\mathbf{E}(z,t)\).  We assume that, for all parts of the sample outside the MCT layer,  \(\mathbf{P}_{\mathrm{e}}(z,t) = 0\) and for all but the MCT and CdTe layers, \(\mathbf{P}_\mathrm{ph}(z,t) = 0\).  All of these materials are semi-insulating semiconductors with fairly large gaps:  1.5 eV for CdTe, 2.3 eV for ZnTe and 1.5 eV for GaAs.  This suggests that electronic contributions to the dynamics of the polarization density are negligible at THz frequencies, justifying our assumption that \(\mathbf{P}_{\mathrm{e}}(z,t) = 0\) in these layers. 
For the CdTe layer, we use $\varepsilon_\infty$=7.61 from Baars and Sorger \cite{Baars1972} and we include in the polarization density \(\mathbf{P}_\mathrm{ph}(z,t)\) only the contribution of a single CdTe-like phonon-polariton with polarization density \(\mathbf{P}_2(z,t)\) and solving Eq. 11 with the initial condition of zero polarizaiton density. 
Our assumption that \(\mathbf{P}_\mathrm{ph}(z,t) = 0\) in the ZnTe buffer layer relies on the fact that the  transversal optical phonon frequency is 5.3 THz in ZnTe~\cite{Schall2001}, which, however, is just outside frequency range of our measurement.  Despite that, since the thickness of this buffer layer is extremely small, we believe that simplifying the response by neglecting this frequency-dependent contribution does not lead to appreciable errors. 
Differently, in GaAs the TO mode frequency is 8.1 THz~\cite{Moore1996} which allows us to neglect frequency-dependent vibrational contributions to the THz optical properties over most of our frequency range.
We, nevertheless, take into account the effect of the phonon-polariton in ZnTe and GaAs in our spectral range. Given our spectral range, we include it as a frequency-independent contribution by substituting what we call $\varepsilon_{\mathrm{{\infty}}}$ with the low-frequency limit \(\varepsilon_{\mathrm{0}}\) of the dielectric function. We use: \(\varepsilon_{\mathrm{0}} = 9.99\) in ZnTe~\cite{Schall2001}, and \(\varepsilon_{\mathrm{0}} = 12.9\) in GaAs~\cite{Moore1996}.

The optical properties of the MCT layer are strongly dependent on the doping \(x\),  which varies over the layer thickness as described by Orlita et al. \cite{Orlita2014} for a similar sample.  The MCT of Orlita et al. can be considered as a sequence of three layers.  The first (adjacent to the buffer layers) has a thickness of about 1.7 \textmu m and has a doping that varies monotonically from $x = 0.31$ to $x = 0.17$.  This is followed by 
a region of about $3.2$ \textmu m thickness with a relatively constant doping of $x=0.17$.  The final layer is about 400 nm thick and has a doping that rapidly increases from $x=0.17$ to $x=0.46$ at the vacuum surface. In our particular sample, grown using the same methods~\cite{Teppe2016}, the relatively constant region has a slightly higher Cd-doping of $x=0.175$, and so we approximate our $x(z)$ by rescaling the composition profile of Orlita at al. to make the doping of this constant-composition region the doping of our sample.

For our model, we make two simplifying assumptions regarding the optical properties in the MCT layer.  The first assumption is that free carriers (and therefore contributions to \(\mathbf{P}_{\mathrm{e}}(z,t)\)) exist only in the central 3.2 \textmu m-thickness later where the doping is constant.  This implies that \(\mathbf{P}_{\mathrm{e}}(z,t) = 0\) in the other portions of the material.  We justify this assumption by noting that according to the empirical band gap expression~\cite{Hansen1982}
\begin{equation}
\mathcal{E}_{\mathrm{g}}(x,T)=-0.302+1.93x+5.35\cdot 10^{-4}T(1-2x)-0.810x^2+0.832x^3,
\end{equation}
where $\mathcal{E}_{\mathrm{g}}$ is in units of eV and $T$ is the temperature in units of K, the gap is a strongly increasing function of $x$ over its domain for temperatures between 10~K and 300~K.  This implies that the carrier concentration in the outer layers of the MCT due to the initial thermal population or carrier generation \cite{Kane1960,Dick1972} is likely much smaller than that of the central region and can be neglected.  Within the central layer we use estimates of the temperature-dependent carrier concentration \(n\), asymptotic velocity 
\(\tilde{c}\), band gap \(\mathcal{E}_{\mathrm{g}}\), effective mass \(m_{\mathrm{o}}^{\mathrm{*}}\) and carrier concentration taken from the data of Teppe et al.~\cite{Teppe2016} \(n_\textrm{lit}\), summarized in Table \ref{tab1}.  

The phonon-polariton terms were kept the same throughout the layer. In MCT, multiple vibrational resonances are present and can generally be divided in HgTe- and CdTe-like bands \cite{Kozyrev1998}. In our simulations, we model each band as a single electric-dipole active resonance, whose parameters vary as a function of temperature.  Thus we have the adjustable fitting parameters $\nu_{\mathrm{TO,1}}$, $\nu_{\mathrm{LO,1}}$ and 
$\gamma_{\mathrm{1}}$ to represent  the HgTe-like modes in Eq. 11 of the main text. This ``single mode'' approximation is fairly good for the HgTe-like modes, since for $x=0.175$ there is a single mode with a much larger oscillator strength. 
For the CdTe-like band we have the fitting parameters $\nu_{\mathrm{TO},2}$, $\nu_{\mathrm{LO},2}$ and $
\gamma_{\mathrm{2}}$.  Unlike the HgTe-modes, the CdTe-like band does not have a single mode with a much stronger oscillator strength, making our approximation here not as good in a general case.
That said, the CdTe-like modes have frequencies at the edge of our investigated spectral window, making a clear distinction between their contributions not possible from the available data.
In our simulations we also use the same parameters $\nu_{\mathrm{TO},2}$, $\nu_{\mathrm{LO},2}$ and $
\gamma_{\mathrm{2}}$ for our treatment of the CdTe buffer layer. In principle, the CdTe-like mode frequency shifts depending on the MCT composition \cite{Kozyrev1998}. However, given that we are not able to spectrally separate such contributions (around 0.3 THz change) from the experimental data in Fig. 1(a), we decided to simplify the model and reduce the number of free parameters by considering only a single common CdTe-like mode.
The phonon parameters are summarized in summarized in Table \ref{tab2}.  

The last remaining parameter for the MCT layer is the high-frequency relative permittivity  $\varepsilon_{\mathrm{{\infty}}}$.
Here we fix this using the composition profile based on the data of Orlita et al.~\cite{Orlita2014} in combination with the empirical relation between $x$ and $\varepsilon_{\mathrm{{\infty}}}$ of Baars and Sorger \cite{Baars1972} to obtain $\varepsilon_{\mathrm{{\infty}}}$ within the film.  Here we use the optical constants measured at 77~K, but these parameters appear to be (up to the experimental resolution) the same at room temperature \cite{Dornhaus1976}.

In our simulations we perform all calculations of the fields over a spatial grid ranging from \(z = z_{\mathrm{-1}}\) to  \(z = z_5\)  where \(z_{\mathrm{-1}} = -13\) \textmu m and \(z_5 = z_{\mathrm{4}} + 52\) \textmu m with step size \(\Delta z\).   
We numerically solve Maxwell’s equations in the time domain using the FDTD method \cite{Yee1966} following the algorithm described by Rumpf \cite{Rumpf2022}, given the initial conditions of zero electric, magnetic and polarization fields in the bulk
and using the displacement field from Eq. 2 of the main text. We adopt perfectly absorbing boundary conditions for the electric and magnetic fields at the boundaries of the simulation grid.

\begin{table}[!h]
\begin{center}
\small
\begin{tabular}{ |c|c|c|c|c|c|c| } 
\hline
$T$ (K) & $n$ (10\textsuperscript{21}m\textsuperscript{-3}) & $\gamma_{\mathrm{P}}/{2\pi}$ (THz) &  $\tilde{c}$ (10\textsuperscript{6} m/s) & $\mathcal{E}_{\mathrm{g}}$ (meV) & $m^{\mathrm{*}}_{\mathrm{o}}$ (10\textsuperscript{-3} $m_{\mathrm{el}}$) & $n_{\mathrm{lit}}$  (10\textsuperscript{21}m\textsuperscript{-3})\\

\hline
13 & 0.6 & 0.8 & 1.06 & 8.6 &0.8& 0.4 \\ 
20 & 0.5 & 0.8 & 1.06 & 9.5 &1.0& 0.4\\ 
40 & 0.5 & 0.7 & 1.05 & 20.8 &1.9& 0.5\\ 
60 & 0.7 & 0.5 & 1.05 & 29.7 &2.8&  1.1\\ 
80 & 1.2 & 0.2 & 1.04 & 38.7 &3.2& 2.1\\ 
100 & 1.8 & 0.2 & 1.04 & 47.2 &3.4& 3.6\\ 
120 & 2.6 & 0.3 & 1.03 & 57.1 &5.2& 6.4\\ 
180 & 6.5 & 0.5 & 1.02 & 83.7 &7.5& 13.0\\ 
240 & 15.1 & 0.8 & 1.00 & 110.6 &10.4& 30.3 \\ 
300 & 25.0 & 0.9 & 0.99 & 137.6 &13.7& 55.3\\ 

\hline

\end{tabular}
\caption{Electronic parameters used in the FDTD simulations. The parameters $\tilde{c}$, $\mathcal{E}_{\mathrm{g}}$, $m^{\mathrm{*}}_{\mathrm{o}}$ and $n_{\mathrm{lit}}$ were estimated from linear interpolation (or extrapolation, beyond 120 K) of the data from Teppe et al.~\cite{Teppe2016}.}
\label{tab1}
\end{center}
\end{table}

\begin{table}[!h]
\begin{center}
\small
\begin{tabular}{ |c|c|c|c|c|c|c|c|c|c|c| } 
\hline
$T$ (K) &  $\nu_{\mathrm{TO,1}}$ (THz) & $\nu_{\mathrm{LO,1}}$ (THz) & $\gamma_{\mathrm{1}}$ (THz) & $\nu_{\mathrm{TO,2}}$ (THz) & $\nu_{\mathrm{LO,2}}$ (THz) & $\gamma_{\mathrm{2}}$ (THz)\\

\hline
13 & 3.6 & 4.4 & 0.8 & 4.45 & 4.65 & 1.0  \\ 
20 & 3.6 & 4.4 & 0.8 & 4.45 & 4.65 & 1.0  \\ 
40 & 3.6 & 4.4 & 0.8 & 4.45 & 4.6 & 1.1 \\ 
60 & 3.6 & 4.4 & 0.8 & 4.4 & 4.6 & 1.1 \\ 
80 & 3.6 & 4.4 & 1.1 & 4.4 & 4.55 & 1.2 \\ 
100 & 3.6 & 4.4 & 1.0 & 4.35 & 4.55 & 1.2 \\ 
120 & 3.6 & 4.3 & 1.7 & 4.35 & 4.55 & 1.2 \\ 
180 & 3.6 & 4.5 & 2.5 & 4.30 & 4.5 & 1.3 \\ 
240 & 3.6 & 4.45 & 2.5 & 4.25 & 4.5 & 1.2 \\ 
300 & 3.6 & 4.42 & 3.5 & 4.25 & 4.45 & 2.1 \\ 

\hline

\end{tabular}
\caption{Phonon parameters used in the FDTD simulations.}
\label{tab2}
\end{center}
\end{table}

\section{2D TDS setup}
\begin{figure}[h]
\includegraphics[width=14 cm]{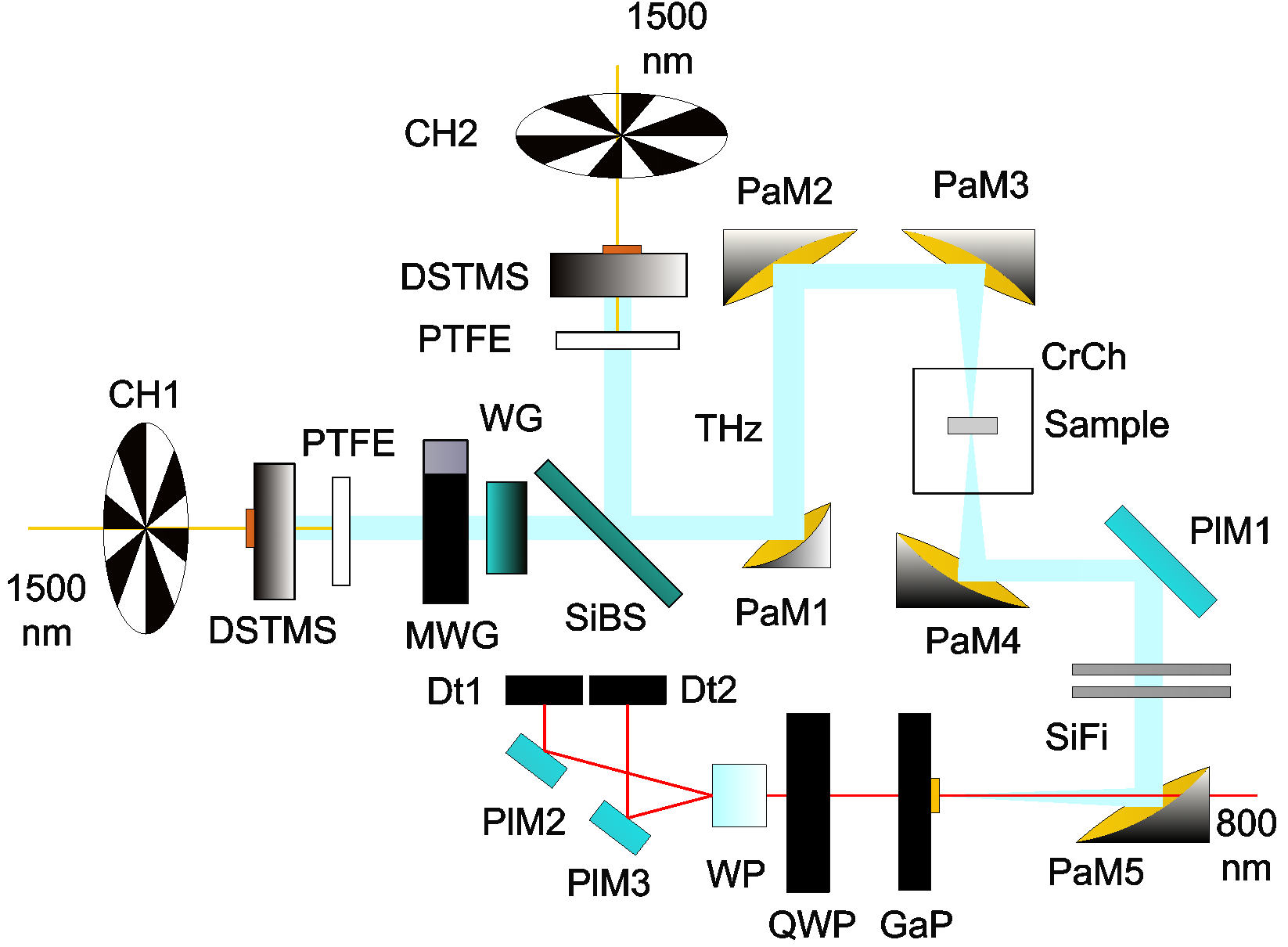}

\caption{Scheme of the 2D-TDS setup. A description is given in the text.} 

 \label{sfig1}
\end{figure}

In Fig. \ref{sfig1}, we schematically present the 2D-TDS setup. In the figure, CH1/CH2 are the two choppers in front of the DSTMS organic crystals generating the $E_{\mathrm{1}}$/$E_{\mathrm{2}}$ THz pulses via optical rectification using the OPA outputs at 1500 nm. To block the residual infrared radiation, two thin polytetrafluoroethylene (PTFE) films (thickness $<$ 100 \textmu{m}) are used.
The $E_{\mathrm{1}}$ field amplitude and direction is regulated by the rotation of a motorized wire grid (MWG) relative to the orientation of a fixed wire-grid polarizer (WG). The $E_{\mathrm{1}}$ beam path merges with the $E_{\mathrm{2}}$ beam at the silicon beamsplitter (SiBS), where they are partially redirected, horizontally polarized, towards a series of parabolic mirrors (PaMs) to first collimate, focus at the sample in the cryostat chamber (CrCh) and recollimate again the THz beams. The transmitted pulses are then redirected through a plane mirror (PM), and focused onto the 300-\textmu{}m GaP crystal. To measure the electric field waveform of the THz pulses in the GaP, we co-propagate a small fraction of the 800 nm femtosecond laser with the THz into the GaP.  The polarization state of the 800 nm pulse is rotated by an amount proportional to the electric field by the Pockels effect.  To measure the changes in polarizaton,  the 800 nm beam passes through a quarter waveplate (QWP) and is then divided into two beams with orthogonal linear polarizations using a Wollaston polarizer (WP).  Two diode detectors (Dt) then measure the intensities of these components.  In the absence of an electric field inside the GaP the diodes measure the same intensities of the 800 nm components;  in the presence of a field the difference is proportional to the electric field at the time when the 800 nm pulse overlaps the THz field in the GaP.

At high electric field amplitudes, the polarization of the 800 nm electro-optic sampling beam can be also influenced by the quadratic Kerr effect.
To minimize this contribution, silicon filters (SiFi) are used to reduce the electric field amplitude to levels where these quadratic contributions are negligible.

\section{1D-TDS at low field}
\begin{figure}[h]
\includegraphics[width=\textwidth]{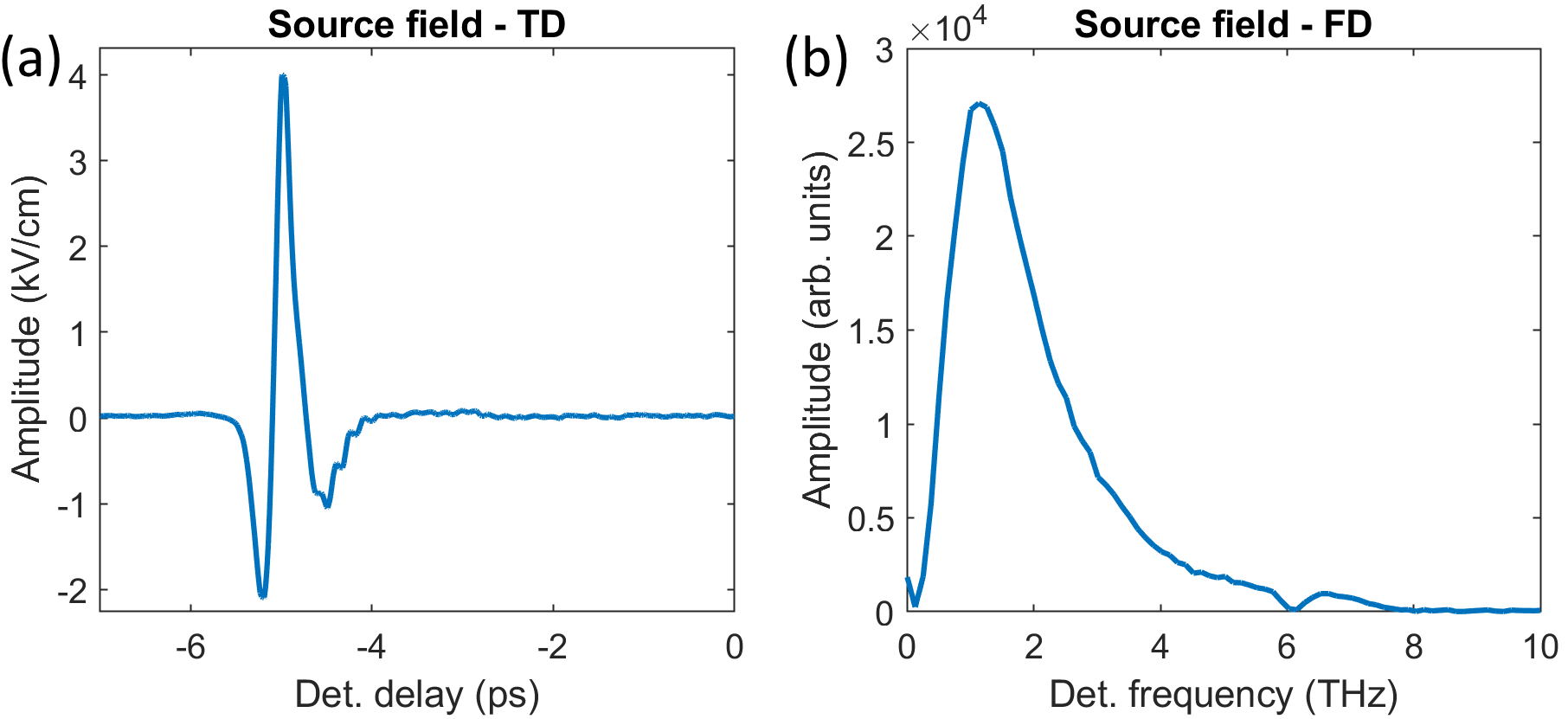}

\caption{Interpolated experimental THz pulse used for the 1D-TDS simulations in the (a) time and (b) frequency domains.} 

 \label{sfig2}
\end{figure}

We record the THz waveform used in the 1D-TDS experiments by sending the beam through a circular hole of the same size as the hole in the support of the sample. A linear interpolation of the waveform is then used as the input for the corresponding FDTD simulations. Fig. \ref{sfig2} shows this waveform in the time and frequency domains. The dip at approximately 6 THz corresponds to absorption from a polytetrafluoroethylene (PTFE) film, used to filter out the residual infrared beam.

\begin{figure}[h]
\includegraphics[width=\textwidth]{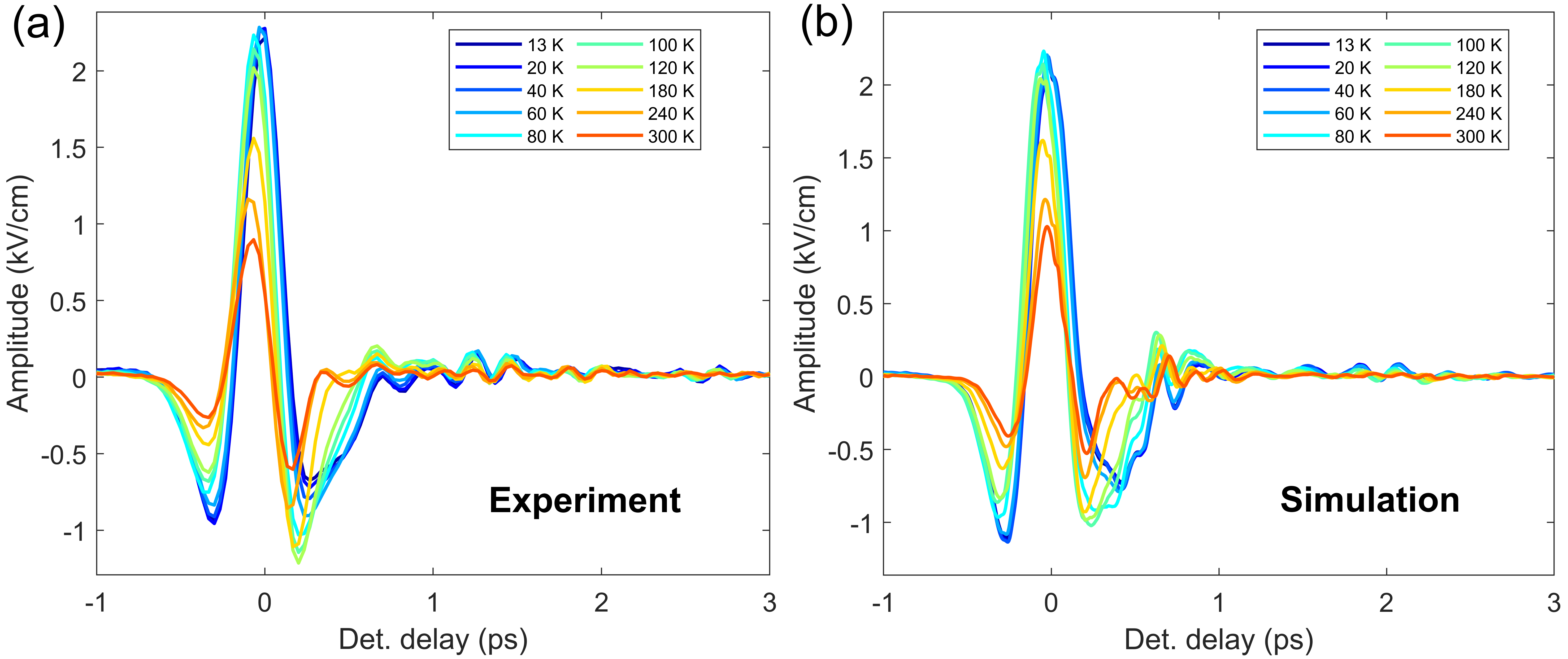}

\caption{(a) Experimental and (b) FDTD-simulated 1D-TDS traces of the transmitted field through the Hg$_{0.825}$Cd$_{0.175}$Te sample using the same input pulse (peak field amplitude $\approx$4 kV/cm). The discrete Fourier transforms of the  traces are reported in Fig. 1 of the main text.} 

 \label{sfig3}
\end{figure}

In Fig. \ref{sfig3} we present the comparison between experimental and simulated FDTD time-domain spectroscopy traces at a low peak field amplitude ($\approx$4 kV/cm) at different temperatures.
The parameters used in the simulations are summarized in Tables \ref{tab1} and \ref{tab2}.

\clearpage
\section{2D - TDS source fields}
\begin{figure}[h]
\includegraphics[width=\textwidth]{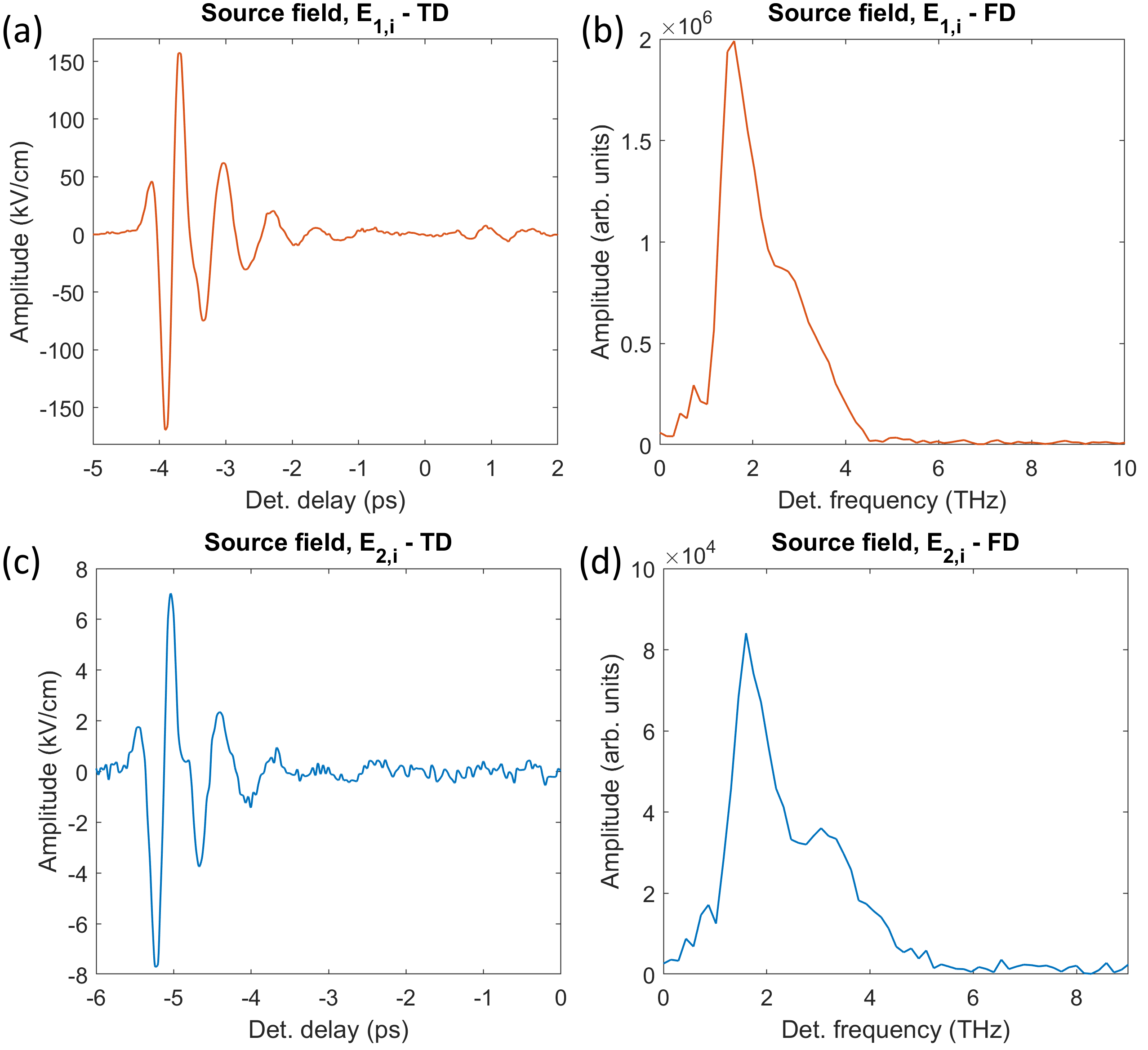}

\caption{Interpolated experimental THz pulses used for the 2D-TDS simulations as the $E_{\mathrm{1,i}}$ and $E_{\mathrm{2,i}}$ pulses in the (a),(c) time and (b),(d) frequency domains.} 

 \label{sfig4}
\end{figure}

Similarly to the 1D-TDS measurements, we record the THz waveforms used in the 2D-TDS experiments by sending the beam through a circular hole of the same size as the one that frames the sample. After linear interpolation, these waveforms are used as the source fields for the corresponding FDTD simulations. We report them in Fig. \ref{sfig4} in the time and frequency domains.

\clearpage


\section{Additional 2D-TDS results}

\begin{figure}[t]
\includegraphics[width=\textwidth]{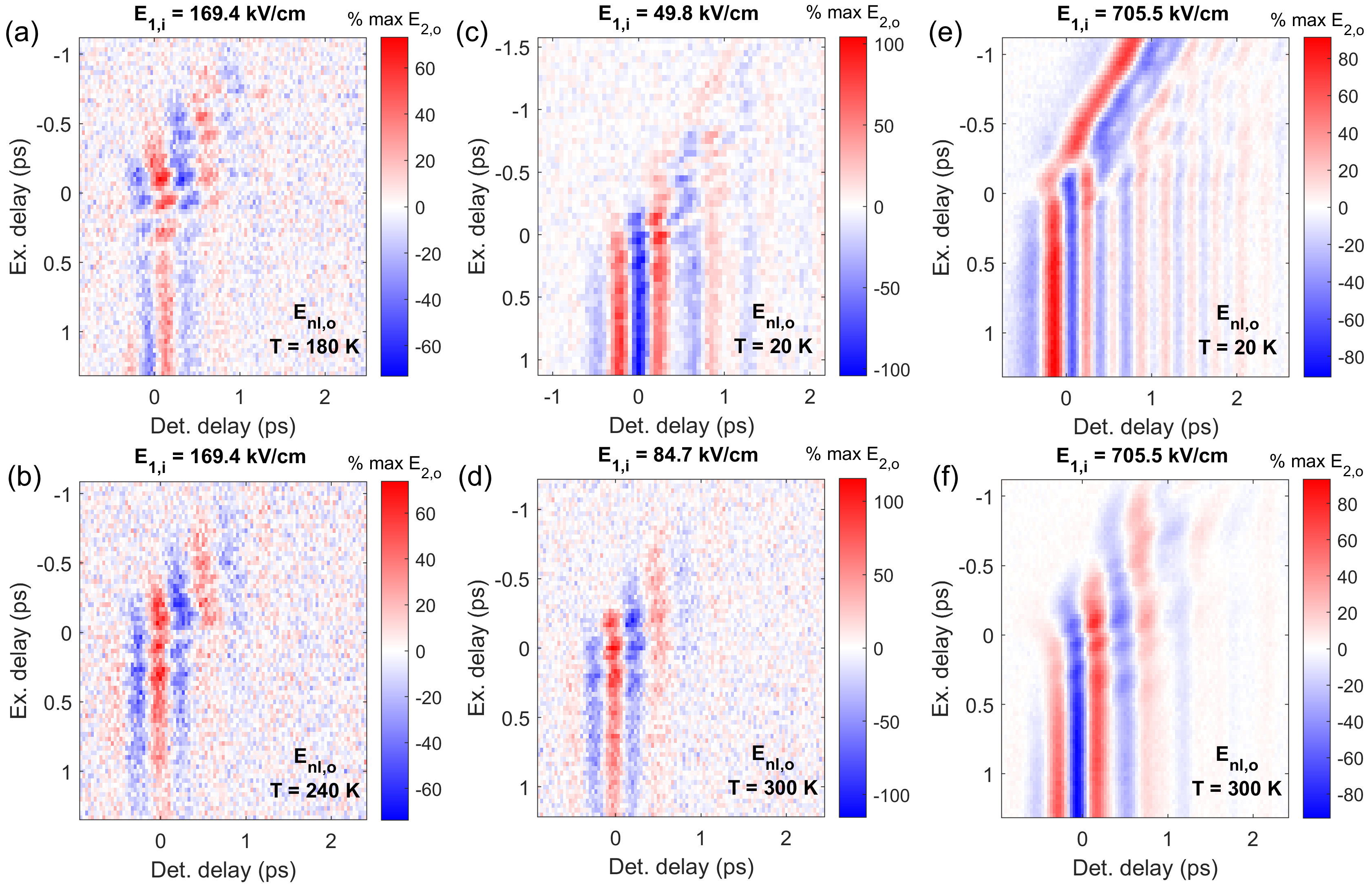}

\caption{Experimental 2D-TDS maps for Hg$_{0.825}$Cd$_{0.175}$Te for $E_{\mathrm{nl,o}}$ at different conditions. (a),(b) Data acquired using peak $E_{\mathrm{1,i}}=169.4$~kV/cm and $E_{\mathrm{2,i}}=7.7$~kV/cm field amplitudes at $T=180$~K and $240$ K. (c) Nonlinear signal at $T=20$~K and peak $E_{\mathrm{1,i}}=49.8$~kV/cm and $E_{\mathrm{2,i}}=7.7$~kV/cm (d) Nonlinear signal at $T=300$ K and peak $E_{\mathrm{1,i}}$=84.7 kV/cm and $E_{\mathrm{2,i}}$=7.7 kV/cm (e),(f) High-field measurements collecting under peak $E_{\mathrm{1,i}}=705.5$~kV/cm and $E_{\mathrm{2,i}}=138.2$~kV/cm field amplitudes at $T=20$~K and $300$~K respectively.}

 \label{sfig5}
\end{figure}

In Fig. \ref{sfig5}(a),(b) we show additional 2D-TDS maps of the nonlinear signal $E_{\mathrm{nl,o}}$ with $T=180$~K and $T=240$~K under the same peak electric field amplitudes as for the data reported in Fig. 2, namely $E_{\mathrm{1,i}}=169.4$~kV/cm and $E_{\mathrm{2,i}}=7.7$~kV/cm. 
The corresponding 2D maps for lower $E_{\mathrm{1,i}}$ amplitudes at $T=20$~K and $T=300$~K, whose constant detection-delay profiles are reported in Fig. 3(a),(b), are shown in Fig. \ref{sfig5}(c),(d).
In Fig. \ref{sfig5}(g),(h) we report the nonlinear effects for the high-field combination $E_{\mathrm{1,i}}$=705.5 kV/cm and $E_{\mathrm{2,i}}$=138.2 kV/cm peak field amplitudes at $T=20$~K and 300~K. In both cases, the nonlinear response corresponds to a decrease in the transmission of the pulses. A proper simulation of these results would require implementation of the full band structure to consider phenomena such as intervalley scattering.

\begin{figure}[t]
\includegraphics[width=\textwidth]{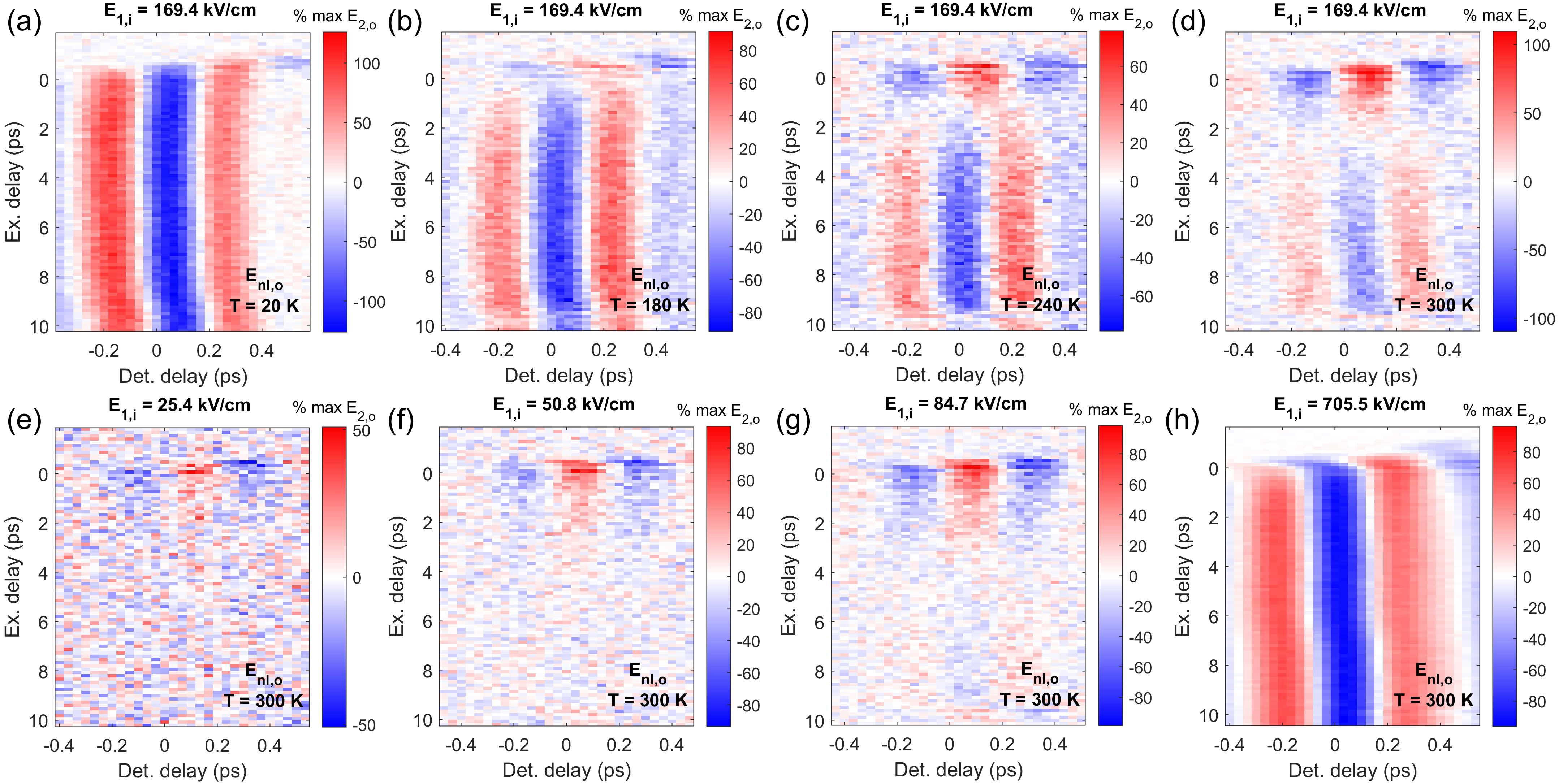}

\caption{Intermediate-delay-range 2D-TDS maps of the nonlinear signal for Hg$_{0.825}$Cd$_{0.175}$Te. (a)-(d) Nonlinear signal $E_{nl}$ at $T= 20$~K, 180~K, 240~K and 300~K temperatures using peak $E_{\mathrm{1,i}}=169.4$~kV/cm and $E_{\mathrm{2,i}}=7.7$~kV/cm field amplitudes (e)-(g) Field dependence of the nonlinear signal at 300 K using peak values $E_{\mathrm{1,i}}= 25.4$~kV/cm, 50.8 kV/cm, 84.7 kV/cm and $E_{\mathrm{2,i}}=7.7$~kV/cm field amplitudes. (h) Nonlinear signal at 300 K using peak $E_{\mathrm{1,i}}= 705.5$~kV/cm and $E_{\mathrm{2,i}}=138.2$~kV/cm field amplitudes.}

 \label{sfig6}
\end{figure}

We next present the dynamics of the nonlinear response for larger values of the excitation delay. To limit the acquisition times, we restricted the range of delay times to the vicinity  of the $E_{\mathrm{2,o}}$ pulse. In Fig. \ref{sfig6}(a)-(d) we show the temperature dependence of the nonlinear effects for $T=20$~K, 180~K, 240~K and 300~K using peak $E_{\mathrm{1,i}}=169.4$~kV/cm and $E_{\mathrm{2,i}}=7.7$~kV/cm. We note that the long-lasting nonlinearity dominant at low temperature, corresponding to a reduction of the transmission, decreases and appears somewhat later as the temperature grows. 
The sharp feature at a excitation delay of approximately 9.5~ps is most likely the consequence of a portion of pulse 1 that is reflected from the GaAs-air and  MCT-air interfaces, effectively creating a delayed copy of pulse 1 that reexcites the sample. At $T=300$~K, the delayed copy induces similar transient nonlinearities as the main part of pulse 1, albeit less pronounced; a similar response was observed for Bi$_{1-x}$Sb$_x$ films \cite{Katayama2018}.
The internal reflection is naturally present also at low temperatures, where it causes an increase in the transmission similar to the high temperature results, differently from the first pump which causes a marked reduction in the transmission. This is linked to the different systems they encounter; in both cases the second pump arrives during an excited transient state with a much larger number of carriers compared to the low-temperature equilibrium, which, in this sense, resembles the high-temperature equilibrium state.

Panels (d)-(h) in Fig. \ref{sfig6} show the long-excitation-delay nonlinear signal at $T=300$~K for some different values of the peak field of pulses 1 and 2. 
We can see here a difference in the field-dependence of the different components of the nonlinear response:  while the rapid and short-lived increase in transmission is already quite visible at low peak fields, the decrease in transmission that usually develops at longer times is measurable only for relatively high peak fields.

\begin{figure}[t]
\includegraphics[width=\textwidth]{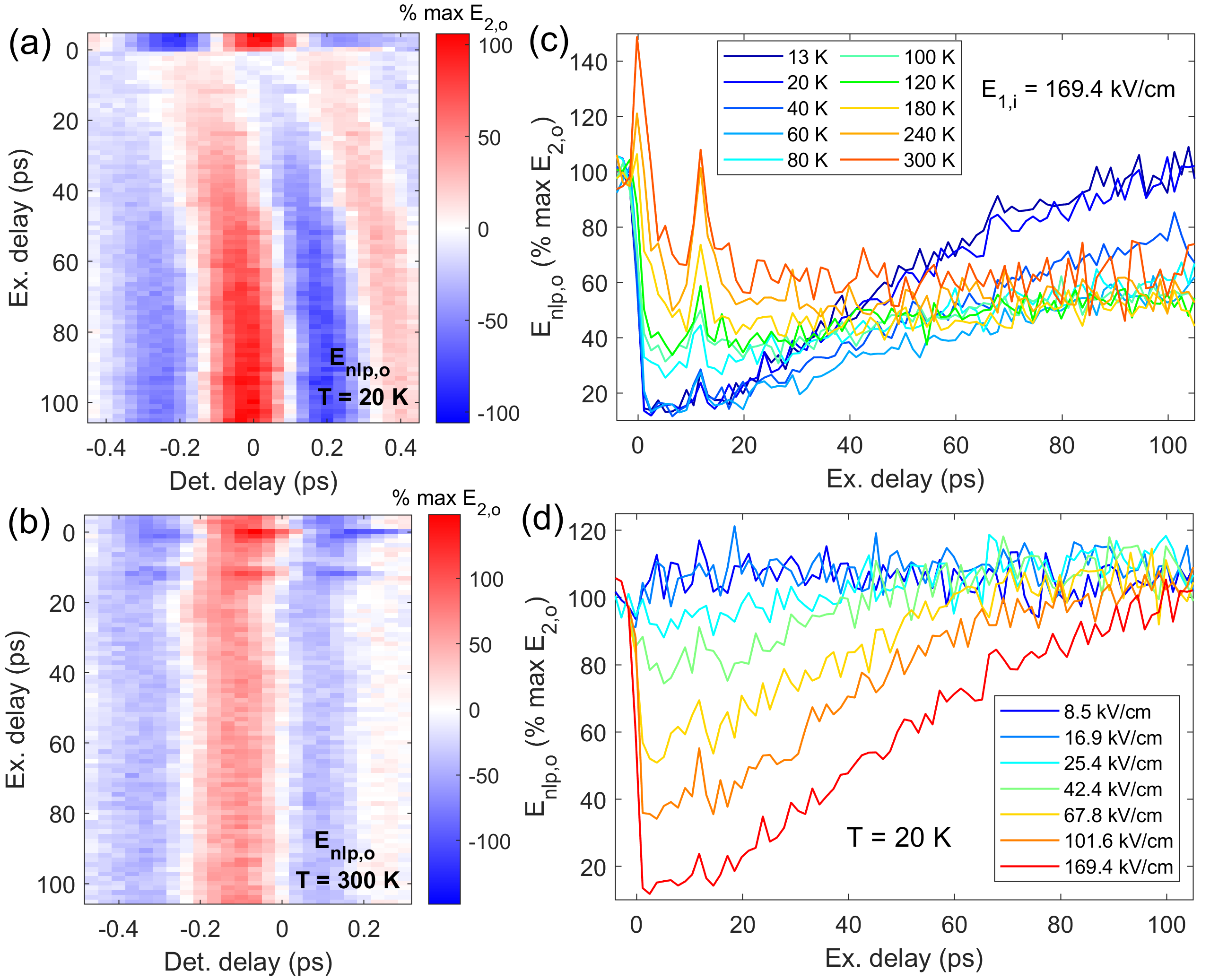}

\caption{Long-delay 2D-TDS maps of the nonlinear signal for Hg$_{0.825}$Cd$_{0.175}$Te. Nonlinear probe $E_{\mathrm{nlp,o}}$ at (a) $T= 20$ K and (b) $300$ K temperatures using peak $E_{\mathrm{1,i}}=169.4$ kV/cm and $E_{\mathrm{2,i}}=7.7$~kV/cm field amplitudes.
(c) Temperature dependence of the time dynamics of the maximum of $E_{\mathrm{nlp,o}}$ using peak $E_{\mathrm{1,i}}=169.4$~kV/cm and $E_{\mathrm{2,i}}=7.7$~kV/cm field amplitudes. (d) Peak $E_{\mathrm{1,i}}$ field dependence of the time dynamics of the maximum of $E_{\mathrm{nlp,o}}$ using a peak field $E_{\mathrm{2,i}}=7.7$ kV/cm at $T=20$~K.}

 \label{sfig7}
\end{figure}

We also measured the nonlinear signal over an excitation delay range of approximately 100 ps. Fig.~\ref{sfig7}(a),(b) show the nonlinear probe $E_{\mathrm{nlp,o}}$ for $T=20$ K and $T=300$~K, respectively, using peak $E_{\mathrm{1,i}}=169.4$~kV/cm and $E_{\mathrm{2,i}}=7.7$~kV/cm.
Here the action of the internal refection as a second pump is evident at around $t_{\mathrm{ex}}\approx$ 10 ps.
Panel (c) shows the time dynamics of the maximum of the nonlinear probe for several temperatures for peak $E_{\mathrm{1,i}}=169.4$~kV/cm and $E_{\mathrm{2,i}}=7.7$~kV/cm.  Panel (d) shows the same but for a fixed temperature $T=20$ K and a variety of values of peak field for $E_{\mathrm{1,i}}$.
While we see that at 20 K all input field strengths result in a nearly complete recovery of the system in 100~ps, the time-dependent changes at higher temperatures are more complex.

First of all, the effects of the main part of pulse 1 and the doubly-reflected copy of pulse 1 are of opposite sign at low temperature, while they both lead to a transient increase of the transmission at higher temperatures. Moreover, the recovery of the equilibrium state takes place with a time scale which grows with the temperature, and is not measurable at room temperature in our time window. The recombination processes are affected not only by temperature-activated contributions, \textit {e.g.}, phonons, but also by the band opening of the sample. 
Furthermore, in panel (a), we observe a lateral shift of the signal not present at high temperature in panels (b). This is a gradual change in the out-of-equilibrium polarization whose detailed description would require the full terahertz trace for each time delay, which is beyond the scope of the work.

\section{Parity of the nonlinear effects}

\begin{figure}[h]
\includegraphics[width=14 cm]{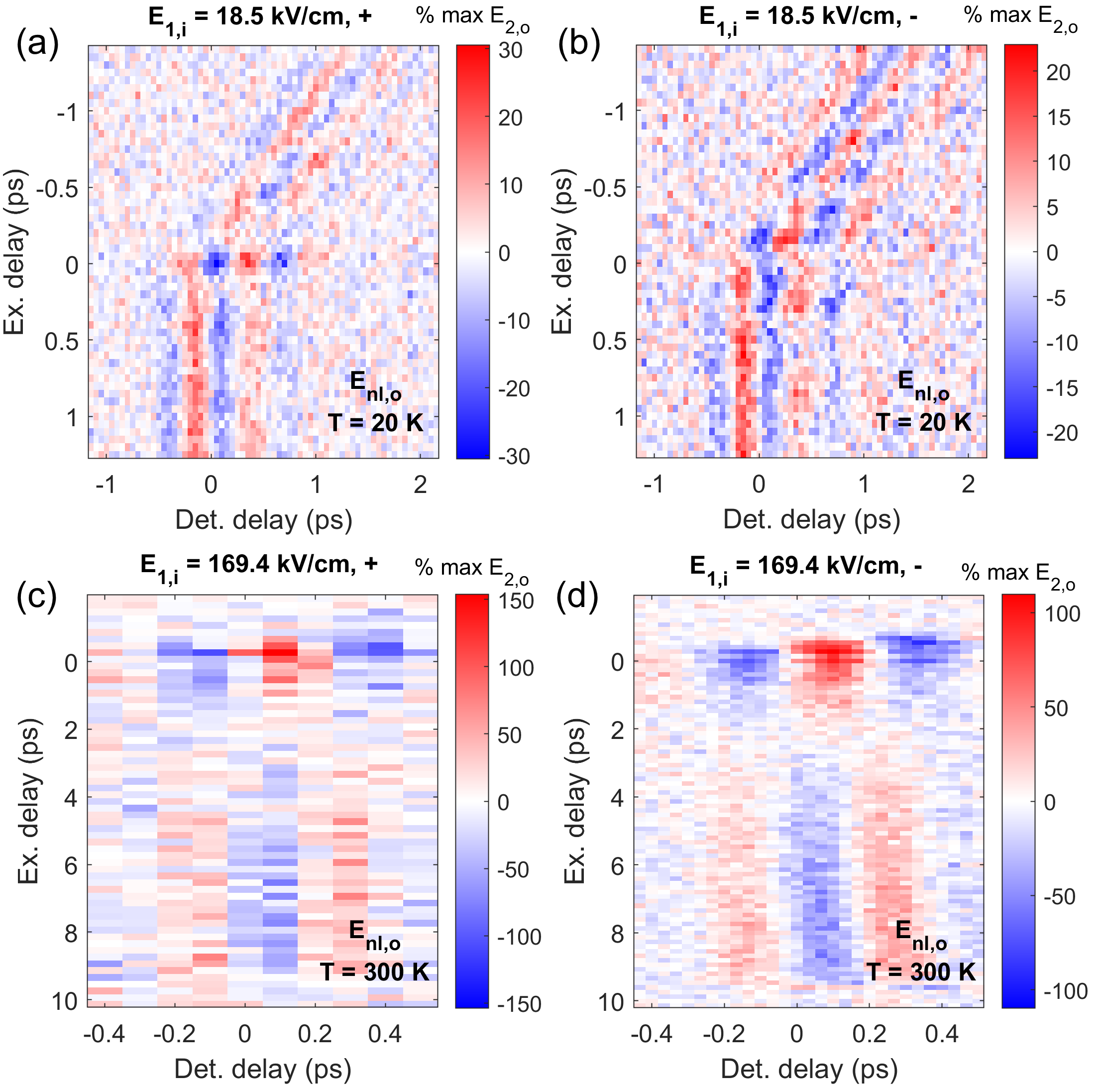}

\caption{2D-TDS maps of the nonlinear signal acquired applying an overall $E_{\mathrm{1,i}}$ sign (+ or -) change at (a),(b) T = 20 K using peak $E_{\mathrm{1,i}}= 18.5$ kV/cm and $E_{\mathrm{2,i}}= 7.7$ kV/cm (c),(d) $T = 300$ K using peak $E_{\mathrm{1,i}}= 169.4$ kV/cm and $E_{\mathrm{2,i}}= 7.7$ kV/cm.} 

 \label{sfig8}
\end{figure}

In order to explore the dependence of the nonlinear signal on parity, we changed the sign of the $E_{\mathrm{1,i}}$ field for measurements at $T=20$ K and $T=300$ K and acquired subsequent 2D $E_{\mathrm{nl,o}}$ maps (Fig. \ref{sfig8}). In the 20 K case (panels (a),(b)), this was achieved by the combined use of wire grids (see previous setup section), while in the 300 K case (panel (c),(d)), we rotated the associated DSTMS crystal by 180 degrees.

At $T=20$ K, the response for excitation delays that are negative or near zero is affected by the $E_{\mathrm{1,i}}$ sign, while the one at more positive excitation delays appears to be unchanged within the noise level.  We believe that this can be understood as follows.
For $t_{\mathrm{ex}}<0$, $E_{\mathrm{2}}$ arrives before $E_{\mathrm{1}}$ and leads to carrier generation which reduces the $E_{\mathrm{1}}$ transmission.  Thus, for $t_{\mathrm{ex}}<0$, the magnitude of $E_{\mathrm{1,o}}$ is reduced, which gives a nonlinear signal odd with respect to $E_{\mathrm{1,i}}$.
Since $E_{\mathrm{2,i}}$ is unchanged when the sign of $E_{\mathrm{1,i}}$ is reversed, the same does not occur for $t_{\mathrm{ex}}>0$. 

At $T=300$ K, the effects around $t_{\mathrm{ex}}\approx0$ are connected to the ballistic motion of the carriers. However, as we saw in the main text, the oscillating pattern in Fig. 2(h),(i) along a detection delay shows modulations always of the same sign, creating
effectively a periodicity at double the $E_{\mathrm{1}}$ frequency of maximum spectral content ($\approx$2$\cdot$1.8=3.6 THz). This results from the field superpositions that determine the nonlinear changes in the field transmission through the electronic polarization and was reproduced by our simulations. Therefore, given the `even' periodicity of the pulse field sign, we expected a similar `even' behaviour when flipping the whole pulse sign. While recorded with coarser time steps and lower integration time than panel (d), the data acquired with the rotated crystal in panel (c) shows the same overall sign pattern of the nonlinear signal.

\section{Estimation of the experimental uncertainties}

There are a few sources of uncertainties in the experiment and simulations. For the experiment, these arise from the electro-optic sampling through the balanced photodetection, and for the simulations some uncertainties stem from the values of the used parameters.

\begin{figure}[h]
\includegraphics[width=14 cm]{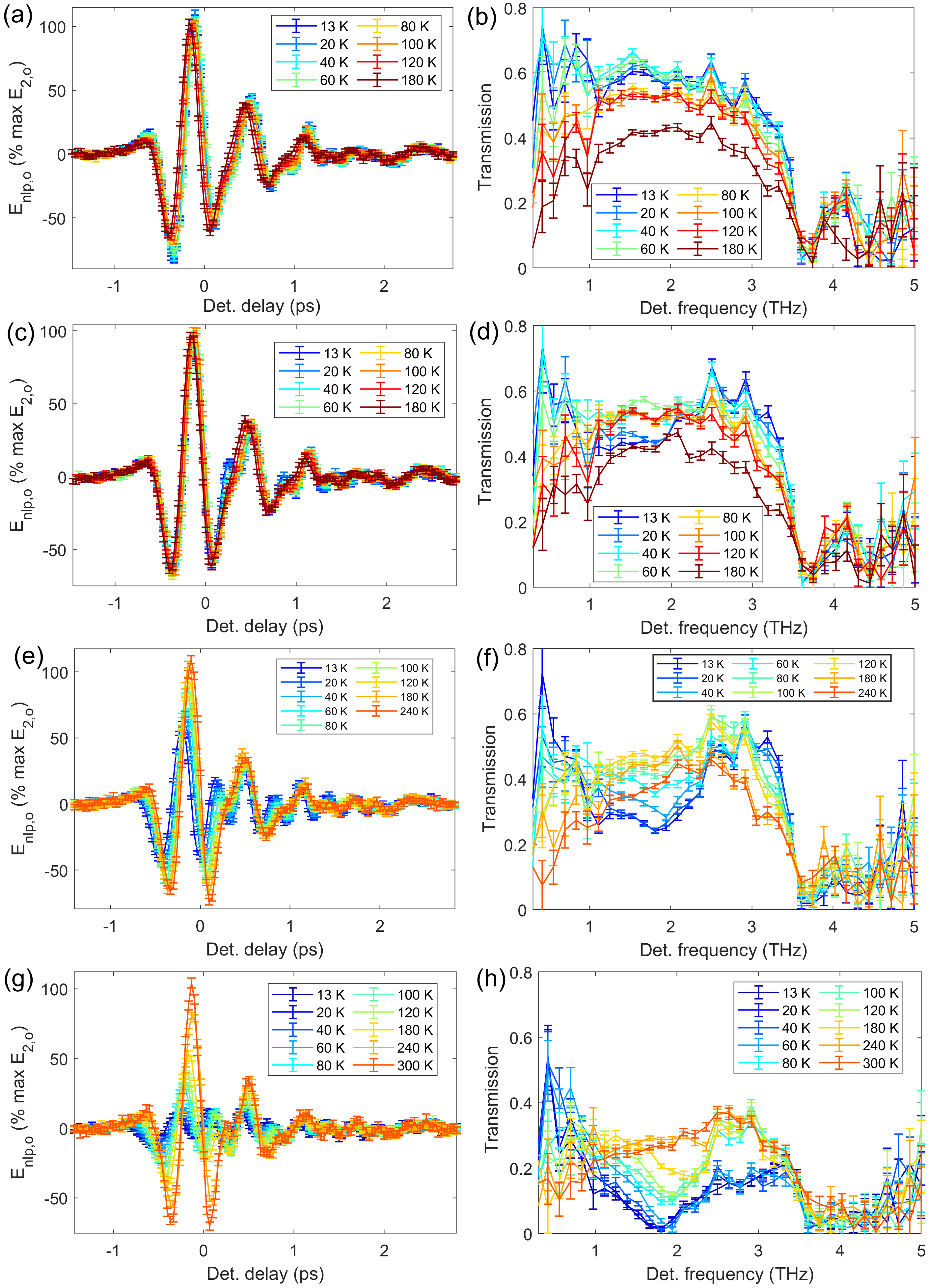}

\caption{Samples of typical data showing experimental uncertainties. Panels (a),(c),(e),(g) show the nonlinear probe $E_{\mathrm{nlp,o}}$ acquired at $t_{\mathrm{ex}}=2.54$~ps (averaged around a $\pm 67$ fs range) using peak fields $E_{\mathrm{1,i}}=169.4$, 50.8, 25.4, 8.5 kV/cm and $E_{\mathrm{2,i}}=7.7$~kV/cm respectively. Panels (b),(d),(f),(h) show the transmission of $E_{\mathrm{2,i}}$ (peak amplitude at 7.7 kV/cm) associated to the data in the corresponding panel on their left.} 

 \label{sfig9}
\end{figure}

The experimental uncertainties can be estimated through repeated measurements of the recorded voltage, assuming a normal distribution of the values. For the time-domain data, the uncertainty for each data point is calculated as the standard deviation of the mean. Derived quantities such as the nonlinear probe $E_{\mathrm{nlp,o}}$ or the nonlinear signal $E_{\mathrm{nl,o}}$ have their uncertainties calculated through error propagation. On average, $E_{\mathrm{nlp,o}}$ is $\approx$3-4 \% with respect to the maximum of $E_{\mathrm{2,o}}$ recorded after the sample, while for $E_{\mathrm{nl,o}}$ is $\approx$5 \%.
The uncertainties in the time domain can be translated in the frequency domain via error propagation \cite{Fornies-Marquina1997}, neglecting numerical uncertainties coming from the discrete Fourier transform algorithm.
Here we report as examples of this analysis the counterparts of the frequency-domain figures (Fig. \ref{sfig9}).

\clearpage

\section{Substrate transmission}
\begin{figure}[h]
\includegraphics[width=10 cm]{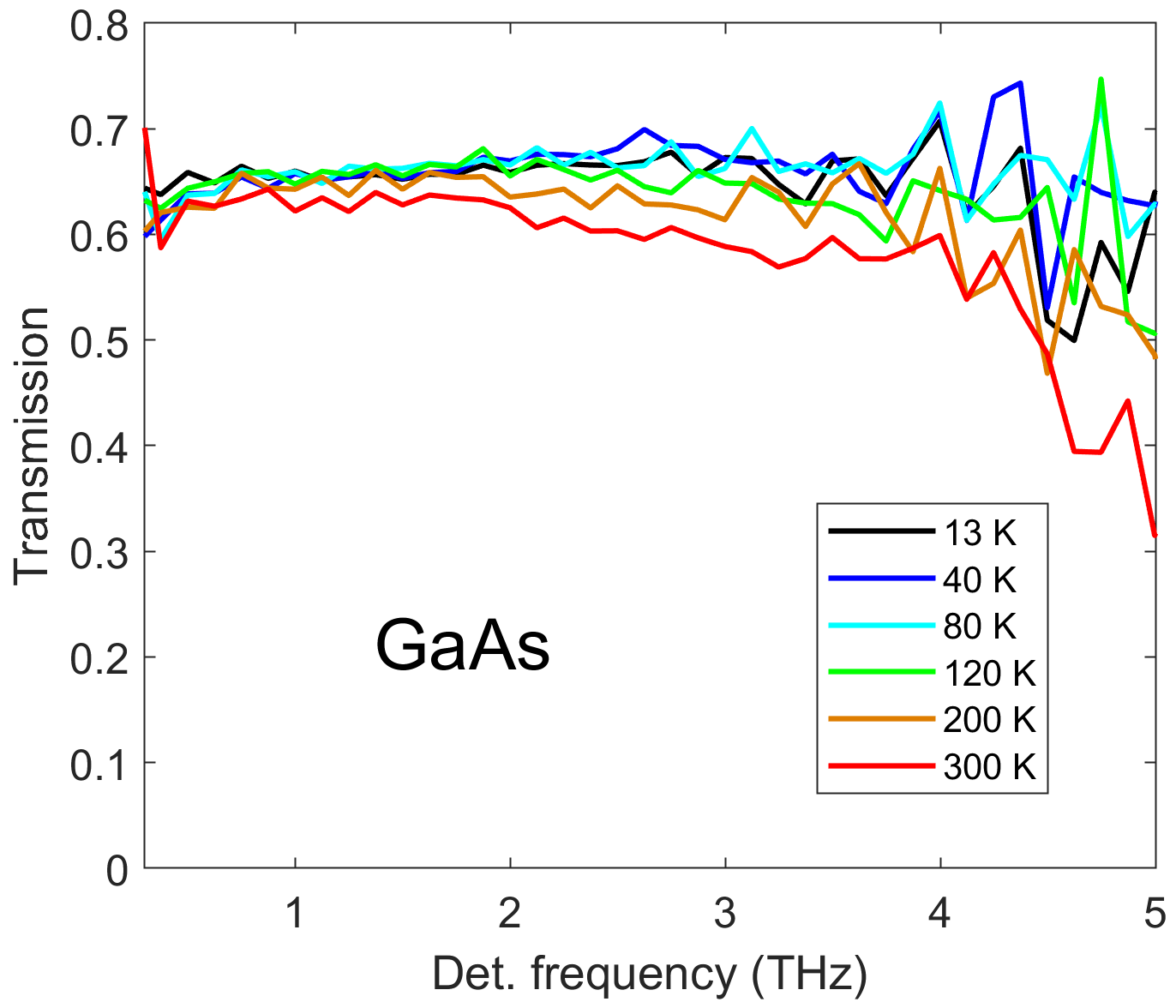}

\caption{Temperature dependence of the GaAs field transmission spectrum for a (110) 500-\textmu{}m-thick sample.} 

 \label{sfig10}
\end{figure}

Using the low-field 1D-TDS we acquired the field transmission of a 500-\textmu{}m-thick undoped semi-insulating GaAs (110) double-polished wafer (Fig. \ref{sfig10}). The transmission is nearly constant over the spectral range where the DSTMS-generated THz field has most of its spectral content (1-4 THz). Varying the temperature in the range 13-300 K, the transmission spectrum shows considerable changes in the transmission amplitude only above 4 THz. Similarly to the MCT sample, rotating the electric field polarization direction of the incident THz field around the normal to the sample surface does not lead to appreciable differences in the transmission spectrum.

\section{Impact ionization simulations}
\begin{figure}[h]
\includegraphics[width=17 cm]{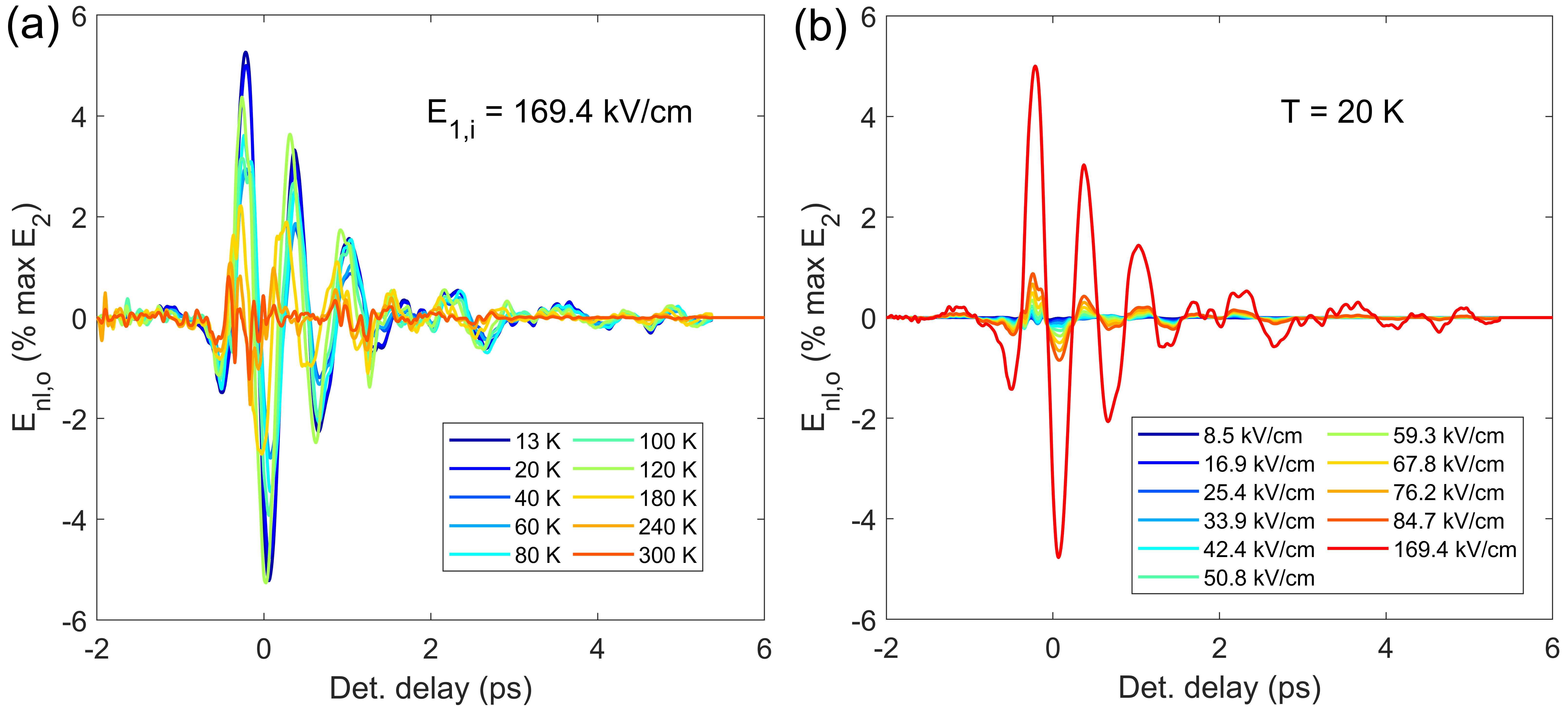}

\caption{FDTD impact ionization simulations using only impact ionization as carrier generation mechanism. (a) Temperature dependence, using peak $E_{\mathrm{1,i}}=169.4$ kV/cm and $E_{\mathrm{2,i}}=7.7$ kV/cm (b) Peak $E_{\mathrm{1,i}}$ field dependence at $T=20$ K with peak field $E_{\mathrm{2,i}}=7.7$ kV/cm.} 

\label{sfig11}
\end{figure}

In this section we report results of the FDTD simulations using only impact ionization as a carrier generation mechanism, applying the model for HgCdTe described by Kinch~\cite{Kinch2004}. The ionization rate is given by Eq. 8 in the main text. The resulting nonlinear signal (Fig. \ref{sfig11}) is much lower compared to the corresponding Zener-only simulations which supports our decision of treating its contribution as minor for the employed THz field amplitudes.

\section{Confinement of the ballistic nonlinear response}
\begin{figure}[h]
\includegraphics[width=\textwidth]{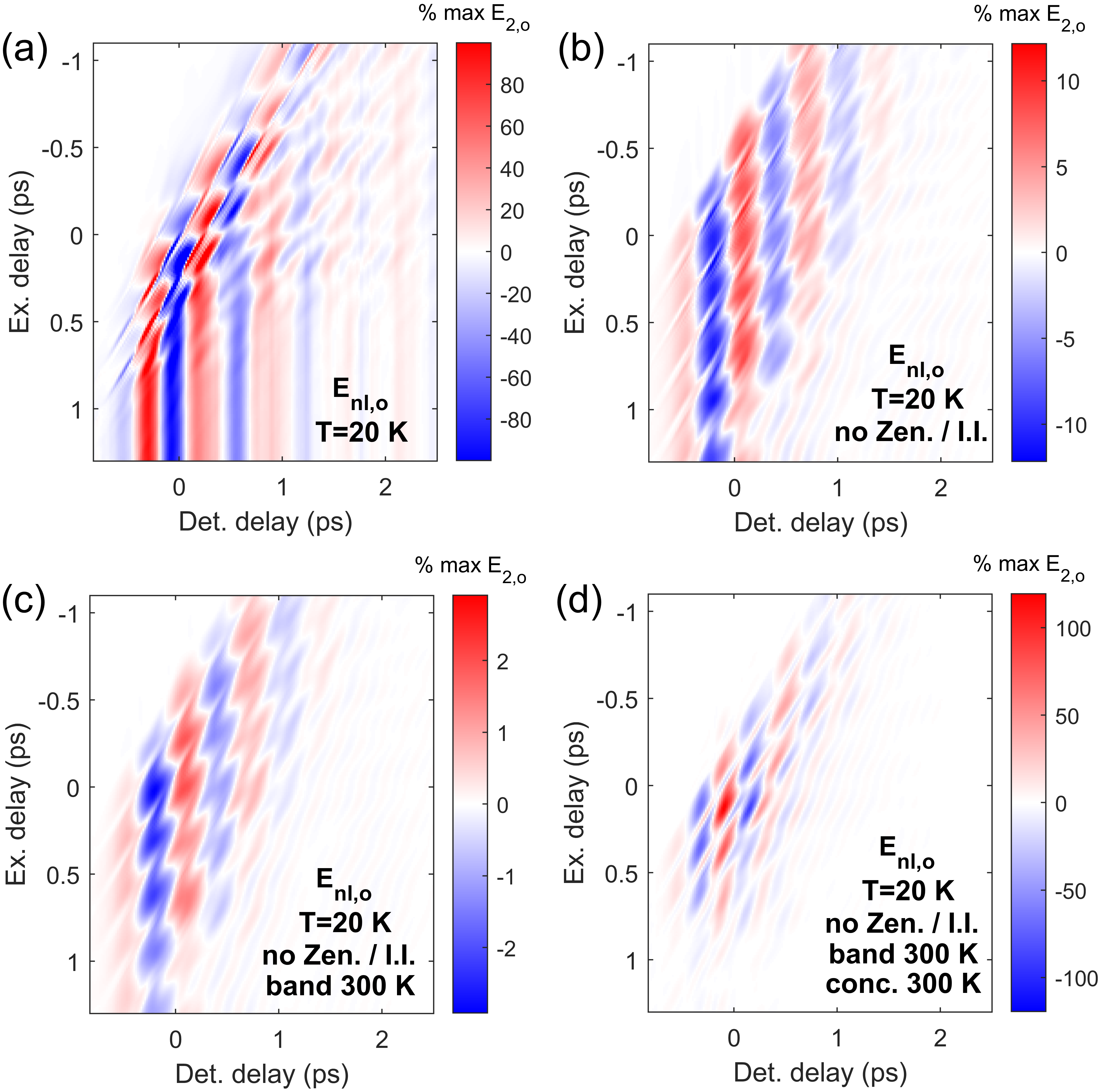}

\caption{2D-TDS plots of the nonlinear signal from FDTD simulations performed using $E_{\mathrm{1,i}}=$ 49.8 kV/cm and $E_{\mathrm{2,i}}=$ 7.7 kV/cm at 20 K. (a) Ballistic + Zener response (b) Ballistic-only response (c) Ballistic-only response using the 300 K band structure (d) Ballistic-only response using the 300 K band structure and carrier concentration.} 

 \label{sfig12}
\end{figure}

In Fig. \ref{sfig12}, we report the results of FDTD simulations for peak fields $E_{\mathrm{1,i}}=$ 49.8 kV/cm and $E_{\mathrm{2,i}}=$ 7.7 kV/cm at 20 K where we explore the dependence on the choice of electronic parameters, and its dependence on the presence of carrier generation due to Zener tunneling.  Panel (a) shows the results using the electronic parameters at 20~K from Table \ref{tab1} to model ballistic dynamics and with Zener tunneling as a carrier generation mechanism. Panel (b) shows the simulation with no Zener tunneling contributions. Here we see that the nonlinear response occurs over a smaller range of excitation delays for a given detection delay.  Panel (c) also shows the simulation without Zener tunneling, but now using the values of $\tilde{c}$ and $m_{\mathrm{o}}^{\mathrm{*}}$ from Table \ref{tab1} for 300 K instead of those for 20 K, resulting in a more parabolic band shape.  Here we see a smaller overall nonlinear signal that is more springily peaked at the time of pulse overlap.  Panel (d) shows the results of the simulation as in panel (c) but also using the higher carrier concentration \(n\) at 300 K.  Here we see a much higher nonlinearity that is strongly peaked at the overlap region.

\begin{figure}[h]
\includegraphics[width=\textwidth]{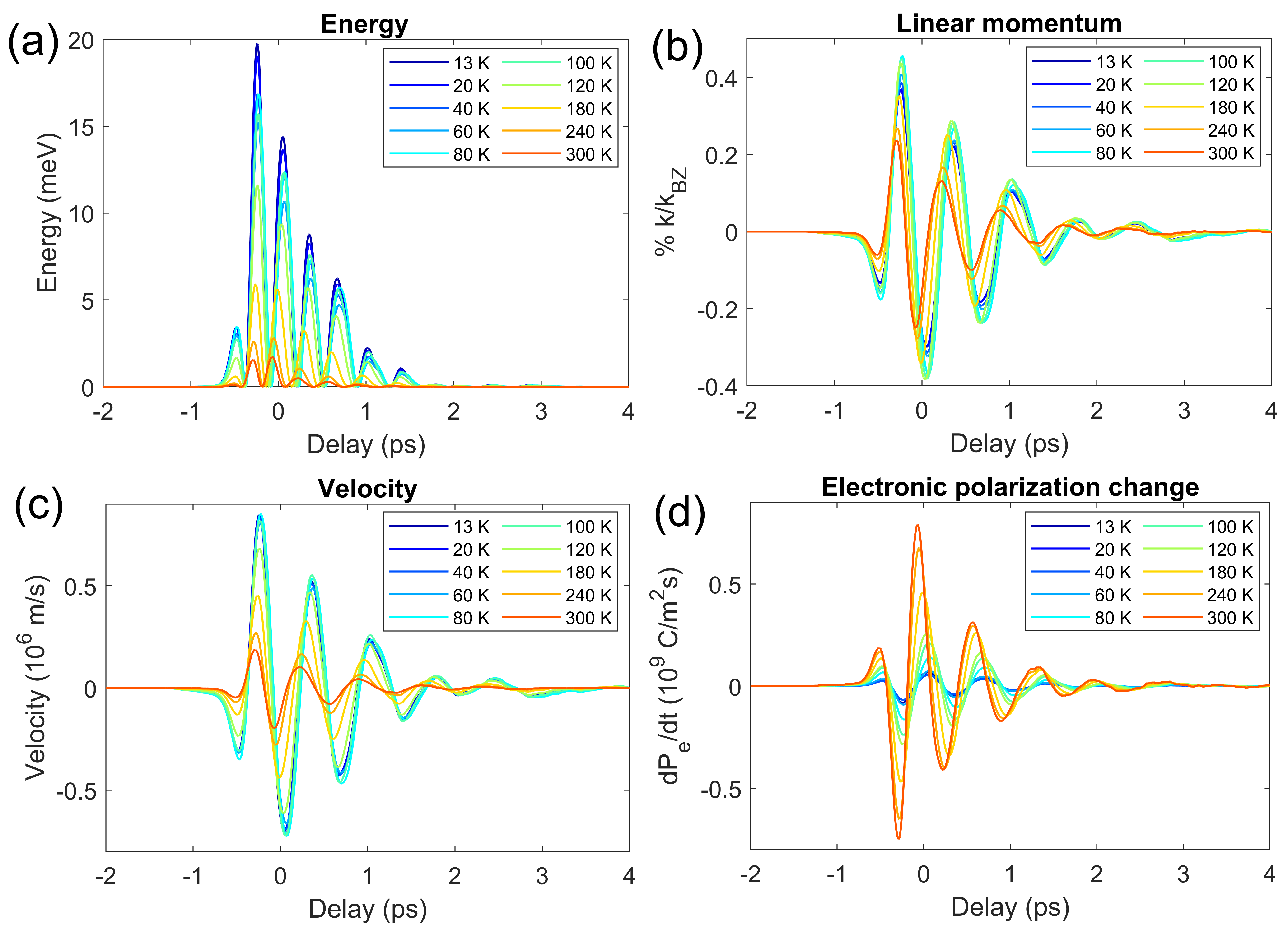}

\caption{Temperature dependence of the ballistic-only electronic dynamics in the first simulation cell of the Hg\textsubscript{0.825}Te\textsubscript{0.175}Te layer using $E_{\mathrm{1,i}}$=8.5 kV/cm. (a) Energy with respect to the bottom of the conduction band (b) Linear momentum (c) Velocity (d) Electronic polarization change.} 

 \label{sfig13}
\end{figure}

Finally, we report in Fig. \ref{sfig13} some ballistic-only simulations performed at different temperatures using as single field $E_{\mathrm{1,i}}$=8.5 kV/cm. As discussed in the main text, the temperature-dependent band shape greatly influences the carrier dynamics at the speed reached by the electrons. While comparable regions of the Brillouin zone are explored in all the cases, the energy value and velocity reached is much higher at low temperature for the same incident field. However, due to the low carrier density, this is not reflected in the electronic polarization change which is larger at higher temperature. 
Moreover, experimentally, the  motion at low temperature is heavily affected by  carrier generation, which reduces the visibility of the ballistic effects in the nonlinear maps with respect to high temperature. 

\section{Carrier generation mechanisms for multicycle pulses}
\begin{figure}[h]
\includegraphics[width=12 cm]{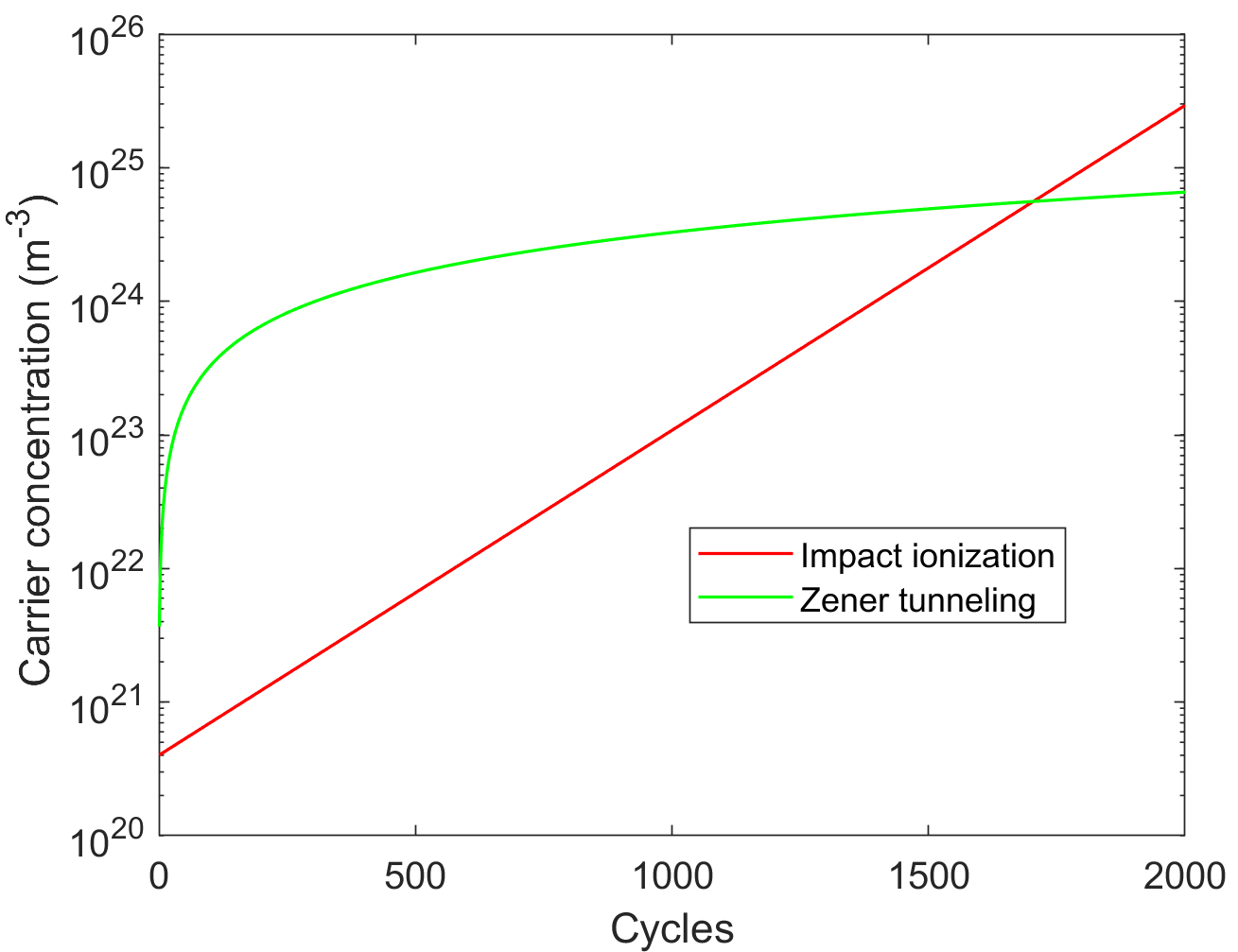}

\caption{Simplified simulation of the carrier generation driven by Zener tunneling and impact ionization mechanisms using a field peak amplitude of 15 kV/cm.} 

\label{sfig14}
\end{figure}

To address the observation of impact ionization characteristics reported by Hubmann et al. regarding the excitation of MCT quantum wells using $\approx$100 ns THz pulses \cite{Hubmann2019}, we performed a simple qualitative simulation of the carrier generation based on the linearization of the carrier increase through the impact ionization expression \cite{Biasco2022,Kinch2004} (Eq. 10 from the main text) together with electronic tunneling though the Zener effect (Eq. 7 from the main text). Assuming monochromatic radiation as in \cite{Hubmann2019} and repeated square-like pulses of 100 fs with electric peak field amplitude of 15 kV/cm, we obtain (Fig. \ref{sfig14}) that after $\approx$1600 cycles the carrier generation from impact ionization should overcome the Zener mechanism, which are much fewer compared to the ones present in a 100-ns pulse. Naturally, this is simply an illustrative simulation which does not take into account saturation effects in the generation, aside from all the approximations described for the FDTD simulations in the main text. Moreover, a more accurate comparison would require the knowledge of the temporal structure of the pulse in \cite{Hubmann2019}. The 100-fs duration is a rough approximation for the time range, within each THz period, over which carrier multiplication mainly occurs.

\clearpage

\bibliography{biblio}